\documentclass[11pt]{book}
\usepackage{kenmax,xspace,epsfig}
\title{\large\sc A Bayesian approach to source separation
\thanks{Presented at the 19th Int. worskhop on Bayesian and 
 Maximum Entropy methods (MaxEnt 1999), Aug. 2-6, 1999, 
 Boise, Idaho, USA}
}
\author{
{\sc Ali Mohammad-Djafari} \\
{\it Laboratoire des Signaux et Syst\`emes (CNRS-SUPELEC-UPS),} \\
{\it Sup\'elec, Plateau de Moulon, 91192 Gif-sur-Yvette, France.} \\
{\it E-mail: djafari@lss.supelec.fr}
}

\def\babs{\begin{abstract}}             \def\eabs{\end{abstract}}
\def\barr{\begin{array}}                \def\earr{\end{array}}
\def\bcc{\begin{center}}                \def\ecc{\end{center}}

\def\bdes{\begin{description}}          \def\edes{\end{description}}
\def\bdoc{\begin{document}}             \def\edoc{\end{document}}
\def\ben{\begin{enumerate}}             \def\een{\end{enumerate}}
\def\beqn{\begin{eqnarray}}             \def\eeqn{\end{eqnarray}}
\def\beqnl#1{\beqn\label{#1}}           \def\eeqnl#1{\label{#1}\eeqn}
\def\beqnx{\begin{eqnarray*}}           \def\eeqnx{\end{eqnarray*}}
\def\bseqn{\begin{subeqnarray}}         \def\eseqn{\end{subeqnarray}}
\def\beq#1\eeq{\begin{equation}#1\end{equation}}
\def\bal#1\eal{\begin{align}#1\end{align}}
\def\balx#1\ealx{\begin{align*}#1\end{align*}}
\def\beqx{$$}                           \def\eeqx{$$}
\def\bfig{\protect\begin{figure}}       \def\efig{\protect\end{figure}}
\def\bfigx{\protect\begin{figure*}}     \def\efigx{\protect\end{figure*}}
\def\bfl{\begin{flushleft}}             \def\efl{\end{flushleft}}
\def\bfr{\begin{flushright}}            \def\efr{\end{flushright}}
\def\bit{\begin{itemize}}               \def\eit{\end{itemize}}
\def\bpic{\begin{picture}}              \def\epic{\end{picture}}
\def\bqu{\begin{quote}}                 \def\equ{\end{quote}}
\def\bqun{\begin{quotation}}            \def\equn{\end{quotation}}
\def\bsl{\begin{slide}}                 \def\esl{\end{slide}}
\def\btabb{\begin{tabbing}}             \def\etabb{\end{tabbing}}
\def\btabl{\begin{table}}               \def\etabl{\end{table}}
\def\btablx{\begin{table*}}             \def\etablx{\end{table*}}
\def\btab{\begin{tabular}}              \def\etab{\end{tabular}}
\def\btabu{\begin{tabular}}             \def\etabu{\end{tabular}}
\def\btabx{\begin{tabular*}}            \def\etabx{\end{tabular*}}
\def\bbib{\begin{thebibliography}{999}} \def\ebib{\end{thebibliography}}
\def\bver{\begin{verbatim}}             \def\ever{\end{verbatim}}

\def\bm#1{\mbox{\boldmath $#1$}}

\def\zerob{{\bm 0}}
\def\oneb{{\bm 1}}

\def\ab{{\bm a}}
\def\bb{{\bm b}}
\def\cb{{\bm c}}
\def\db{{\bm d}}
\def\eb{{\bm e}}
\def\fb{{\bm f}}
\def\gb{{\bm g}}
\def\hb{{\bm h}}
\def\ib{{\bm i}}
\def\jb{{\bm j}}
\def\kb{{\bm k}}
\def\lb{{\bm l}}
\def\mb{{\bm m}}
\def\nb{{\bm n}}
\def\ob{{\bm o}}
\def\pb{{\bm p}}
\def\qb{{\bm q}}
\def\rb{{\bm r}}
\def\sb{{\bm s}}
\def\tb{{\bm t}}
\def\ub{{\bm u}}
\def\vb{{\bm v}}
\def\wb{{\bm w}}
\def\xb{{\bm x}}
\def\yb{{\bm y}}
\def\zb{{\bm z}}

\def\Ab{{\bm A}}
\def\Bb{{\bm B}}
\def\Cb{{\bm C}}
\def\Db{{\bm D}}
\def\Eb{{\bm E}}
\def\Fb{{\bm F}}
\def\Gb{{\bm G}}
\def\Hb{{\bm H}}
\def\Ib{{\bm I}}
\def\Jb{{\bm J}}
\def\Kb{{\bm K}}
\def\Lb{{\bm L}}
\def\Mb{{\bm M}}
\def\Nb{{\bm N}}
\def\Ob{{\bm O}}
\def\Pb{{\bm P}}
\def\Qb{{\bm Q}}
\def\Rb{{\bm R}}
\def\Sb{{\bm S}}
\def\Tb{{\bm T}}
\def\Ub{{\bm U}}
\def\Vb{{\bm V}}
\def\Wb{{\bm W}}
\def\Xb{{\bm X}}
\def\Yb{{\bm Y}}
\def\Zb{{\bm Z}}

\def\alphab{\bm{\alpha}}
\def\betab{\bm{\beta}}
\def\deltab{\bm{\delta}}
\def\epsilonb{\bm{\epsilon}}
\def\gammab{\bm{\gamma}}
\def\omegab{\bm{\omega}}
\def\thetab{\bm{\theta}}
\def\xib{\bm{\xi}}
\def\lambdab{\bm{\lambda}}
\def\taub{\bm{\tau}}
\def\phib{\bm{\phi}}
\def\mub{\bm{\mu}}
\def\psib{\bm{\psi}}
\def\chib{\bm{\chi}}
\def\sigmab{\bm{\sigma}}

\def\Deltab{\bm{\Delta}}
\def\Lambdab{\bm{\Lambda}}
\def\Phib{\bm{\Phi}}
\def\Psib{\bm{\Psi}}
\def\Sigmab{\bm{\Sigma}}

\def\Ac{{\cal A}}
\def\Bc{{\cal B}}
\def\Cc{{\cal C}}
\def\Dc{{\cal D}}
\def\Ec{{\cal E}}
\def\Fc{{\cal F}}
\def\Gc{{\cal G}}
\def\Hc{{\cal H}}
\def\Ic{{\cal I}}
\def\Jc{{\cal J}}
\def\Kc{{\cal K}}
\def\Lc{{\cal L}}
\def\Mc{{\cal M}}
\def\Nc{{\cal N}}
\def\Oc{{\cal O}}
\def\Pc{{\cal P}}
\def\Qc{{\cal Q}}
\def\Rc{{\cal R}}
\def\Sc{{\cal S}}
\def\Tc{{\cal T}}
\def\Uc{{\cal U}}
\def\Vc{{\cal V}}
\def\Wc{{\cal W}}
\def\Xc{{\cal X}}
\def\Yc{{\cal Y}}
\def\Zc{{\cal Z}}

\def\wt#1{\widetilde{#1}}
\def\wh#1{\widehat{#1}}
\def\xh{\widehat{x}}
\def\thetah{\widehat{\theta}}
\def\betah{\widehat{\beta}}

\def\xbh{\widehat{\xb}}
\def\thetabh{\widehat{\thetab}}
\def\betabh{\widehat{\betab}}

\def\xbhk{\widehat{\xb}^{k}}
\def\thetahk{\widehat{\theta}^{k}}
\def\betahk{\widehat{\beta}^{k}}

\def\xbhkp{\widehat{\xb}^{k+1}}
\def\thetahkp{\widehat{\theta}^{k+1}}
\def\betahkp{\widehat{\beta}^{k+1}}

\def\thetabhk{\widehat{\thetab}^{k}}
\def\betabhk{\widehat{\betab}^{k}}

\def\thetabhkp{\widehat{\thetab}^{k+1}}
\def\betabhkp{\widehat{\betab}^{k+1}}

\def\thetamin{\theta_{\mbox{\tiny min}}}
\def\thetamax{\theta_{\mbox{\tiny max}}}
\def\betamin{\beta_{\mbox{\tiny min}}}
\def\betamax{\beta_{\mbox{\tiny max}}}

\def\ra{\rightarrow}
\def\la{\leftarrow}
\def\da{\downarrow}
\def\ua{\uparrow}

\def\Ra{\Rightarrow}
\def\La{\Leftarrow}
\def\Da{\Downarrow}
\def\Ua{\Uparrow}

\def\lra{\longrightarrow}
\def\lla{\longleftarrow}
\def\Lra{\Longrightarrow}
\def\Lla{\longleftarrow}

\def\lrarr{\leftrightarrow}
\def\Lrarr{\Leftrightarrow}
\def\udarr{\updownarrow}
\def\Uparr{\Updownarrow}

\def\d#1{\,\mbox{d}#1}
\def\dxdy{\d{x}\d{y}}
\def\dwxdwy{\d{\omega_x}\d{\omega_y}}
\def\dxdydz{\d{x}\d{y}\d{z}}

\def\disp#1{{\displaystyle #1}}
\def\diag#1{\mbox{diag}\left\{#1\right\}}

\def\Prob#1{\mbox{Pr}\left\{#1\right\}}
\def\var#1{\mbox{Var}\left\{#1\right\}}
\def\cov#1{\mbox{Cov}\left\{#1\right\}}
\def\corr#1{\mbox{Corr}\left\{#1\right\}}
\def\trace#1{\mbox{Tr}\left\{#1\right\}}
\def\rang#1{\mbox{rang}\left\{#1\right\}}
\def\det#1{\mbox{d\'et}\left\{#1\right\}}

\def\cosf{\cos \phi}
\def\sinf{\sin \phi}
\def\cost{\cos \theta}
\def\sint{\sin \theta}

\def\sgn{\mbox{sgn}}
\def\sinc{\mbox{sinc}}
\def\rect{\mbox{rect}}
\def\sincf#1{\mbox{sinc}\left(#1\right)}
\def\rectf#1{\mbox{rect}\left(#1\right)}
\def\trif#1{\mbox{tri}\left(#1\right)}
\def\xvec#1#2#3{\left\{#1_#2,\ldots,#1_#3\right\}}

\def\vx{\left[x_1,\ldots, x_n\right]^t}
\def\vz{\left[z_1,\ldots, z_n\right]^t}
\def\vw{\left[\omega_1,\ldots, \omega_n\right]^t}
\def\vxi{\left[\xi_1,\ldots, \xi_n\right]^t}
\def\iii{\int_{-\infty}^{+\infty}}
\def\izi{\int_{0}^{\infty}}
\def\izpi{\int_{0}^{\pi}}
\def\izdpi{\int_{0}^{2\pi}}
\def\intd{\int\kern-.8em\int}
\def\intt{\int\kern-.8em\int\kern-.8em\int}
\def\intg{\int\kern-1.1em\int}
\def\sumd{\mathop{\sum\sum}}

\def\sumi{\sum_{i=1}^{M}}
\def\sumj{\sum_{i=1}^{N}}
\def\sumk{\sum_{k=1}^{K}}
\def\sumn{\sum_{n=1}^{N}}
\def\summ{\sum_{m=1}^{M}}

\def\TA#1{{\cal A}\left\{ {#1} \right\}}
\def\TH#1{{\cal H}\left\{ {#1} \right\}}
\def\TP#1{{\cal P}\left\{ {#1} \right\}}
\def\TR#1{{\cal R}\left\{ {#1} \right\}}
\def\TRa#1{{\cal R}^{\dag}\left\{ {#1} \right\}}
\def\BR#1{{\cal B}\left\{ {#1} \right\}}
\def\TF#1{{\cal F}\left\{ {#1} \right\}}
\def\TFI#1{{\cal F}^{-1}\left\{ {#1} \right\}}
\def\TFn#1#2{{\cal F}_{#1}\left\{ {#2} \right\}}
\def\TFnI#1#2{{\cal F}_{#1}^{-1}\left\{ {#2} \right\}}
\def\Im#1{{\cal I}\mbox{m}\left(#1\right)}
\def\Ker#1{{\cal K}\mbox{er}\left(#1\right)}
\def\Imag#1{\mbox{Im}\left(#1\right)}
\def\Re#1{\mbox{Re}\left(#1\right)}
\def\expf#1{\exp\left[ {#1} \right]}

\def\dfdx#1#2{{\mbox{d} {#1}\over{\mbox{d} {#2}}}}
\def\dfdxd#1#2{{\mbox{d}^2 {#1}\over{\mbox{d} {#2}^2}}}
\def\dfdxt#1#2{{\mbox{d}^3 {#1}\over{\mbox{d} {#2}^3}}}
\def\dfdxn#1#2{{\mbox{d}^n {#1}\over{\mbox{d} {#2}^n}}}
\def\dfdxk#1#2{{\mbox{d}^k {#1}\over{\mbox{d} {#2}^k}}}

\def\dpdx#1#2{{{\partial {#1}\over \partial {#2}}}}
\def\dpdxd#1#2{{{\partial^2 {#1}}\over{\partial {#2}^2}}}
\def\dpdxdy#1#2#3{{{\partial ^2 {#1}}\over{\partial {#2} \partial {#3}}}}

\def\arg{\mbox{arg}}
\def\argmins#1#2{\mbox{arg}\min_{#1}\left\{{#2}\right\}}
\def\argmaxs#1#2{\mbox{arg}\max_{#1}\left\{{#2}\right\}}
\def\argmin#1#2{\mathop{\mbox{arg}\min}_{#1}\left\{{#2}\right\}}
\def\argmax#1#2{\mathop{\mbox{arg}\max}_{#1}\left\{{#2}\right\}}

\def\esp#1{\mbox{E}\left\{ #1 \right\}}
\def\espx#1#2{\mbox{E}_{#1}\left\{ #2 \right\}}

\def\wth#1{\widehat{\widetilde{\phantom{#1}}}\!\!\!\! #1}

\def\lrf{L_{r,\phi}}
\def\fw{\widehat{f}(\omegab)}
\def\fthwxi{\wth{f}(\Omega,\xib)}
\def\fthwfi{\wth{f}(\Omega,\phi)}
\def\ftrfi{\widetilde{f}(r,\phi)}

\def\fwxwy{\widehat{f}(\omega_x, \omega_y)}
\def\wxpwy{(\omega_x \, x + \omega_y \, y)}

\def\wtx{\omegab^t \cdot \xb}
\def\ejwtx{\exp\left[j \omegab^t \cdot \xb\right]}
\def\xitx{\xib^t \cdot \xb}

\def\ftrxi{\widetilde{f}(r,\xib)}
\def\ent{-\int p(x) \, \ln p(x) \d{x}}

\def\mean#1{\left< #1 \right>}
\def\slnhn{\sum_{n=1}^N \lambda_n h_n(\rb)}
\def\slngn{\sum_{n=1}^N \lambda_n g_n(\rb)}
\def\smngn{\sum_{n=1}^N \mu_n g_n(\rb)}
\def\slmhm{\sum_{m=1}^N \lambda_m h_m(\rb)}
\def\vlambda{\bm{\lambda} = [\lambda_1,\ldots,\lambda_n]}

\def\apriori{{\em a priori} }
\def\aposteriori{{\em a posteriori} }

\def\titre#1{\bcc{\Large\bf #1}\ecc}

\def\AMD{Ali Mohammad--Djafari}
\def\LSSa{Laboratoire des Signaux et Syst\`emes 
(CNRS--ESE--UPS) \\ 
\'Ecole Sup\'erieure d'\'Electricit\'e \\ 
Plateau de Moulon, 91192 Gif sur Yvette Cedex, France.}

\def\ME{maximum entropy}
\def\pdf{probability distribution function}
\def\lm{Lagrange multipliers}
\def\fix#1{\phi _#1(x)}
\def\fin{\fix n}
\def\fik{\fix k}
\def\fiz{\fix 0}
\def\sfinz{\sum_{n=0}^N \lambda_n \, \fin}
\def\sfinu{\sum_{n=1}^N \lambda_n \, \fin}
\def\bl{\bm{\lambda}}
\def\bd{\bm{\delta}}
\def\blz{\bl ^0}
\def\gnl{G _n(\bl)}
\def\gnlz{G _n(\blz)}
\def\un{n=1,\dots, N}
\def\nn{n=0,\dots, N}

\def\finn{\fin , \nn}
\def\esfinz{\exp\,\left[ -\sfinz \right] }
\def\esfinu{\exp\,\left[ -\sfinu \right] }
\def\esxm{\exp\,\left[ -\sum_{m=0}^N \lambda_m \, x^m \right] }
\def\efin{\esp \fin }
\def\zl{Z(\bl)}
\def\finxi{\phi _n(x_i)}
\def\snfinxi{\sum_{n=1}^N \lambda_n \finxi}
\def\esnfinxi{\exp \left[ - \snfinxi \right]}
\def\smfinxi{\sum_{i=1}^M \finxi}

\def\ejnw{\exp \left( -j n \omega_0 x \right) }
\def\eejnw{\mbox{E} \left\lbrace \ejnw \right\rbrace}

\def\signed#1{{\unskip\nobreak\hfil\penalty50\hskip2em\mbox{}
\nobreak\hfil\tt#1\parfillskip=0pt \finalhyphendemerits=0 \par}}

\def\uncatcodespecials{\def\do##1{\catcode`##1=12 }\dospecials}
\def\listing#1{\par\begingroup\setupverbatim\input#1 \endgroup}
\newcount\lineno
\def\setupverbatim{\tt \lineno=0
 \obeylines \uncatcodespecials \obeyspaces
 \everypar{\advance\lineno by1 \llap{\sevenrm\the\lineno\ \ }}}
{\obeyspaces\global\let =\ }

\def\defined{\stackrel{\mbox{def}}{=}}
\def\str{\stackrel}

\def\ER{\mbox{I\kern-.25em R}}
\def\EC{\mbox{C\kern-.8em C}}
\def\EZ{\mbox{Z\kern-.55em Z}}
\def\EN{\mbox{N\kern-.8em N}}

\def\singles{
 \abovedisplayskip 12pt plus 3pt minus 9pt
 \belowdisplayskip 12pt plus 3pt minus 9pt
 \abovedisplayshortskip 0pt plus 3pt
 \belowdisplayshortskip 7pt plus 3pt minus 4pt
 \baselineskip 14.4pt
 \lineskip 1pt
 \lineskiplimit 0pt}
\def\oneandhalf{
 \abovedisplayskip 18pt plus 3pt minus 9pt
 \belowdisplayskip 18pt plus 3pt minus 9pt
 \abovedisplayshortskip 0pt plus 3pt
 \belowdisplayshortskip 9.333pt plus 3pt
 \baselineskip 20pt
 \lineskip 2pt
 \lineskiplimit 1pt}

\def\double{
 \abovedisplayskip 24pt plus 3pt minus 9pt
 \belowdisplayskip 24pt plus 3pt minus 9pt
 \abovedisplayshortskip 0pt plus 3pt
 \belowdisplayshortskip 12pt plus 3pt
 \baselineskip 27pt
 \lineskip 3pt
 \lineskiplimit 2pt}

\def\dadb{\d{\alpha}\d{\beta}}

\def\ffbox#1{\fbox{\mbox{\vbox{#1}}}}

\def\rot{\mbox{rot}}
\def\case#1#2#3#4{
    \left\{
           \begin{array}{ll}
            {\displaystyle #1} & {\displaystyle #2} \cr 
            {\displaystyle #3} & {\displaystyle #4}
           \end{array}
    \right. }

\def\beqnarr#1&#2&#3\\#4&#5&#6\eeqnarr{
    \left\{
           \begin{array}{lcl}
            {\displaystyle #1} & #2 & {\displaystyle #3} \\ 
            {\displaystyle #4} & #5 & {\displaystyle #6} 
           \end{array}
    \right. }

\def\pyx{p(\yb|\xb)}
\def\pxy{p(\xb|\yb)}

\def\ie{{\em i.e.}}
\def\unsdpi{\left(\frac{1}{2\pi}\right)}
\def\unspi{\left(\frac{1}{\pi}\right)}
\def\up{\uppercase}
\def\zjm{z_{j-1}}
\def\zjp{z_{j+1}}
\def\fxyp{f(x,y)=\left\{
\barr{ll} 1 & (x,y)\in P\\ 0 & (x,y)\not\in P\earr
\right.}

\def\rem#1{}

\def\homedir{/users/bonobo/djafari/}
\def\bibdir{\homedir Tex/Inputs/}
\def\Eps{Eps/}

\def\sumi{\sum_{i=1}^{M} }
\def\sumj{\sum_{j=1}^{N} }
\def\lambdah{\wh{\lambda}}
\def\lambdabh{\wh{\lambdab}}

\def\det#1{\hbox{det}\left(#1\right)}
\def\defined{\,\shortstack{$\triangle$\\ =}\,}
\def\sign{\hbox{sign}}

\def\ss{source separation }
\def\pdf{probability density function }
\def\mm{mixing matrix }
\def\sm{separating matrix }

\def\sbh{\wh{\sb}}
\def\abh{\wh{\ab}}
\def\Sbh{\wh{\Sb}}
\def\Abh{\wh{\Ab}}
\def\Bbh{\wh{\Bb}}

\def\Lambdab{\bm{\Lambda}}
\def\xbh{\widehat{\xb}}

\def\pxta{p\left( \xb(1),\ldots,\xb(T) | \Ab \right)}

\def\pxas{p\left( \xb_{1..T} | \Ab, \sb_{1..T} \right)}
\def\pasx{p\left( \Ab,\sb_{1..T} | \xb_{1..T} \right)}

\def\pax{p\left( \Ab | \xb_{1..T} \right)}
\def\psx{p\left( \sb | \xb_{1..T} \right)}
\def\pbx{p\left( \Bb | \xb_{1..T} \right)}
\def\pxb{p\left( \xb_{1..T} | \Bb \right)}

\def\ria{r_i\left( [\Ab^{-1} \xb]_i \right)}
\def\rib{r_i\left( [\Bb \xb]_i \right)}
\def\riat{r_i\left( [\Ab^{-1} \xb(t)]_i \right)}
\def\ribt{r_i\left( [\Bb \xb(t)]_i \right)}

\def\pia{p_i\left( [\Ab^{-1} \xb]_i \right)}
\def\pib{p_i\left( [\Bb \xb]_i \right)}
\def\piat{p_i\left( [\Ab^{-1} \xb(t)]_i \right)}
\def\pibt{p_i\left( [\Bb \xb(t)]_i \right)}

\def\zmat{\left[\matrix{0}\right]}
\def\det#1{\hbox{det}(#1)}

\begin{document}
\maketitle
\begin{abstract}
Source separation is one of the signal processing's main emerging domain. 
Many techniques such as maximum likelihood (ML), Infomax, cumulant 
matching, estimating function, etc. have been used to address this 
difficult problem. 
Unfortunately, up to now, many of these methods could not account 
completely for noise on the data, for different number of sources 
and sensors, 
for lack of spatial independence and for time correlation of the sources. 
Recently, the Bayesian approach has been used to push farther these 
limitations of the conventional methods. 
This paper proposes a unifying approach to source separation based on 
the Bayesian estimation. We first show that this approach gives the 
possibility to explain easily the major known techniques in sources 
separation as special cases. Then we propose new methods based on maximum 
{\em a posteriori~} (MAP) estimation, either to estimate directly the sources, 
or the mixing matrices or even both. 
\end{abstract}

\section{Introduction}

The simplest model for a source separation is 
\beq
\xb(t)=\Ab \, \sb(t), 
\eeq
where $\Ab$ is a mixing matrix, $\sb(t)$ is a vector of sources and 
$\xb(t)$ a vector of independent measurements. The main task is then 
to recover $\sb(t)$, but one may instead be interested in recovering a 
separating matrix $\Bb$ such that $\sbh(t)=\Bb \, \xb(t)$. When $\Ab$ 
is invertible, it is natural to assume that $\Bb=\Ab^{-1}$ or 
$\Bb=\Sigmab\, \Lambdab\, \Ab^{-1}$ where $\Sigmab$ is a permutation 
matrix and $\Lambdab$ a diagonal scaling matrix. 

Many \ss algorithms have been recently proposed based on 
likelihood \cite{Ziskind88,Stoica96,Wax91,InfoMaxML,ica99:lacoume,ica99:oja,ica99:macleod,ica99:bermond}, 
contrast function \cite{Comon1991,Jutten1991,ica99:moreau1,ica99:comon}, 
estimating function \cite{CL-easi,SOBI-SP,semipara,ProcIEEE}, 
information theory \cite{Bell_Sejnowski,InfoMaxML,ica99:alphey,ica99:pham}, 
and more generally on principle component analysis (PCA) 
\cite{kn:Tipping99}, Independent factor analysis (IFA) \cite{kn:Attias,kn:Press82,kn:PS89}
and independent component analysis (ICA) \cite{ProcIEEE,JADE:NC,Iscas96-algebra}. 
All these methods assume 
that the mixing matrix $\Ab$ is invertible and mainly search for a 
separating matrix $\Bb$ such that the components of 
$\yb(t)=\Bb\, \xb(t)$ be independent. 
This means that all these 
methods implicitly assume that the sources $\sb(t)$ are independent. 
This may not be the case in some applications. 
However, the main differences between 
these methods are in the way they try to insure this independence. 

\bit
\item Maximum likelihood (ML) techniques use directly the 
independence property by assuming
\beq \label{model1}
p(\sb)=\prod_i p_i (s_i)
\eeq
and as a result 
\beq
p(\xb | \Ab) = \frac{1}{\det{\Ab}} \prod_i \pia,  
\eeq
or equivalently 
\beq
p(\xb | \Bb) = |\det{\Bb}| \prod_i \pib 
= |\det{\Bb}| \prod_i p_i(y_i),  
\eeq
where $y_i=[\Ab^{-1} \xb]_i=[\Bb\xb]_i$ and 
where $p(\sb)$ is the \pdf of the source vector $\sb$. 
The ML estimate of the separating matrix is defined as 
\beqn
\Bbh &=& \argmax{\Bb}{\log p(\xb | \Bb) } \nonumber\\ 
         &=& \argmax{\Bb}{\sum_i r_i(y_i) + \log |\det{\Bb}|} 
             \quad \hbox{with}\quad r_i(y_i)=\log p_i(y_i). \qquad ~ 
\eeqn
A great number of algorithms have been proposed to perform this 
optimization \cite{Bell_Sejnowski,InfoMaxML}. 

\item Infomax techniques use the entropy of $\yb=\Bb\xb$ as a 
measure of independence \cite{Cichocki96,kn:Roberts98,Lee99b,Lee99c,Lee99d}: 
\beq
S=-\sum_i p_i(y_i) \log p_i(y_i). 
\eeq
Thus $S$ is a function of 
the \sm $\Bb$ and one tries to optimize $S$ with respect to $\Bb$. 

\item M-estimation techniques define an estimate for the \sm  
$\Bb$ such that 
\beq
\frac{1}{T} \sum_{t} \Hb \left( \yb(t) \right) = \zmat 
\quad \hbox{with}\quad \yb(t)=\Bb \xb(t), 
\eeq
where it is assumed to have $T$ independent observations 
$\left\{ \xb(1), \ldots, \xb(T) \right\}$ 
and where $\Hb$ is an appropriately defined matrix valued function. 
$\zmat$ represents a matrix whose elements are all equal to zero. 
We can note that M-estimate methods generalize the ML estimation 
method since the latter can be obtained by taking 
\beq
\Hb \left( \yb \right) = \dpdx{\log p(\yb)}{\Bb}.
\eeq

\item Contrast function minimization techniques are based on the 
optimization of a contrast function 
$c(\yb)=c\left( \Bb \xb \right)$ which takes its extremal value 
when $\Bb$ is a \sm \cite{Comon1991,Jutten1991}. 
Typical examples are the contrast functions 
measuring, in some way, the independence of the components of $\yb$, 
sometimes subject to the constraint that $\yb(t)$ be spatially white
\beq
\frac{1}{T} \sum_{t} \yb(t)  \, \yb^{\dag}(t) = \Ib. 
\eeq

\item Higher order statistics (HOS) techniques try to insure the 
independence of the components of $\yb$ by minimizing, under the 
whiteness constraint, a contrast function related to the statistics of 
the order greater that two such as the cumulants 
\cite{Cardoso89,Cardoso99a,Cardoso99b}. 
\eit

The main limitations of these techniques are the following:
\bit
\item None of these techniques consider the possible errors on the 
model or the measurement (sensor) noises; 

\item All these methods assume that the \mm $\Ab$ is invertible and 
cannot account for the cases in which $\Ab$ is rectangular (number 
of sensors different from the number of sources). 

\item All these methods assume that the sources are independent. 
Some assume the sources to be also temporally white. 
\eit

Recently, a few works using the Bayesian approach have been presented   
to push farther the limits of these methods 
\cite{kn:Rajan97,kn:Knuth98,kn:Knuth98b,kn:LeePress,kn:Roberts98,kn:Roberts98,kn:Knuth99,Lee99a}. 

In the following, we first present the basics of the Bayesian 
approach, then we show how some of the preceeding techniques can be 
obtained as special cases, and finally, we propose new ideas 
to account for spatial correlation between neighbor sources 
or time correlation of the sources. 

\section{Bayesian approach}

The main idea in the Bayesian approach is to use not only the 
likelihood \\ 
$\pxta$ but also some prior knowledge about the sources 
$\sb$ and the \mm $\Ab$ through the assignment of prior probabilies 
$p(\sb)$ and $p(\Ab)$. Then, noting 
\(
\xb_{1..T}=\{\xb(1),\ldots,\xb(T)\}
\) 
and 
\(
\sb_{1..T}=\{\sb(1),\ldots,\sb(T)\}
\) 
and using these direct probability laws we determine the posterior law 
\beq
\log \pasx = \log \pxas + \log p(\Ab) + T \log p(\sb) + cte, 
\eeq
where we assumed the independence of the sources $\sb$ and the \mm 
$\Ab$. 

From this posterior probability law we can deduce any inference 
about $\Ab$ and $\sb$. For example, we can estimate both $\Ab$ and 
$\sb$ by a joint maximum {\em a posteriori~} (JMAP) criterion using an 
alternate maximization algorithm. We can also focus on the estimation of 
the \mm $\Ab$ by marginalizing this posterior law with respect to 
$\sb$ to obtain $p(\Ab|\xb_{1..T})$ and use the resulting 
MAP criterion to estimate $\Ab$. 
Finally, we can integrate $\Ab$ from the joint law to 
obtain $p(\sb_{1..T}|\xb_{1..T})$ and estimate the sources from this 
marginal posterior law. 

Now, before going further in details of these three methods, we are going to 
illustrate some special cases which result in some classical 
techniques. 

\bigskip
\subsection{Exact invertible model and independent sources}

If we assume that the model $\xb=\Ab\,\sb$ is exact and that there is 
not any measurement noise and that the \mm $\Ab$ is invertible and 
well conditioned, then 
we can only look for a \sm $\Bb=\Ab^{-1}$. Indeed, as in conventional 
methods, if we assume that the sources $\sb$ are independent, we have 
the following relations: 
\beq
p(\sb)=\prod_i p_i(s_i)
\eeq
and so  
\beq
p(\xb | \Bb) = |\det{\Bb}| \prod_i \pib, 
\eeq
where $p_i(s_i)$ is the \pdf of the source component $i$. 
Using these relations, and noting by $\yb(t)= \Bb \,\xb(t)$ we have 

\beq
\log \pbx = \log \pxb + \log p(\Bb) + cte, 
\eeq
where 
\[
\log \pxb = T \log |\det{\Bb}| + \sum_t \sum_i \log p_i\left(y_i(t)\right)
\]
and $p(\Bb)$ is a probability distribution on the \sm $\Bb$. 
Here we assume that we can assign a probability law $p(\Ab)$ to the 
\mm $\Ab$ or equivalently $p(\Bb)$ to the \sm $\Bb$ 
to translate any prior knowledge we have about (or we wish to 
impose to) them. 
For example, we may know (or assume) that the \mm is such that 
\beq
\| \Ab \|^{2} \defined \sum_k\sum_l |a_{k,l}|^{2} \le \epsilon, 
\eeq
for some $\epsilon$; 
or we may wish that the \sm $\Bb$ be such that its determinant 
$|\det{\Bb}| \not= 0$ and not very far from one. 
In the first case we can choose 
\beq \label{pa1}
p(\Ab) \propto \expf{ -\frac{1}{2\sigma_a^2} \| \Ab \|^{2} } 
=\expf{ -\frac{1}{2\sigma_a^2} \sum_k\sum_l a_{k,l}^{2} }
\eeq
and in the second case 
\beq
p(\Bb) \propto |\det{\Bb}|. 
\eeq
Some other possibilities are:

\beq \label{pa2}
p(\Ab) \propto \expf{ -\frac{1}{2\sigma_a^2} \| \Ib - \Ab \|^{2} } 
=\expf{ -\frac{1}{2\sigma_a^2} 
\left[\sum_k(1-a_{k,k})^{2} + \sum_{l\not= k} a_{k,l}^{2} 
\right]}
\eeq
which tries to impose $|a_{k,l}| \simeq 1,\, k=l$ and 
$|a_{k,l}| \simeq 0,\, k\not= l$;

\beq \label{pa3}
p(\Ab) \propto \expf{ -\frac{1}{2\sigma_a^2} \| \Ib - \Ab\Ab^t \|^{2} } 
=\expf{ -\frac{1}{2\sigma_a^2} 
\left[\sum_k(1-\|a_{k,*}\|^2)^{2} -\sum_{l\not= k} [\Ab\Ab^t]_{k,l}^{2} 
\right]}
\eeq
when the number of sources is less than the number of the 
sensors; and 
\beq \label{pa4}
p(\Ab) \propto \expf{ -\frac{1}{2\sigma_a^2} \| \Ib - \Ab^t\Ab \|^{2} } 
=\expf{ -\frac{1}{2\sigma_a^2} 
\left[\sum_l (1-\|a_{*,l}\|^2)^{2} -\sum_{k\not= l} [\Ab^t\Ab]_{k,l}^{2} 
\right]}
\eeq
when the number of sources is greater than the number of the 
sensores. 
These two last expressions have been proposed and used by 
Knuth~\cite{kn:Knuth98}. 
Other choices based on prior knowledge of the geometrical positions of the 
sources and receivers and knowledge of the signal propagation law for 
an acoustical application have been used by 
\cite{kn:Knuth98,kn:Knuth99}. 

Now, if we consider the MAP estimation, the MAP criterion to 
optimize becomes 
\beq
J(\Bb)= \log \pbx 
= T \log|\det{\Bb}| + \sum_{t}\sum_i \log p_i\left(y_i(t)\right) 
+ \log p(\Bb) + cte. 
\eeq
Searching now for the MAP solution, the necessary condition is
\beq \label{grad1}
\dpdx{J(\Bb)}{\Bb}=\zmat \lra 
-\sum_{t} \Hb \left( \yb(t) \right) =\zmat, 
\eeq
where $\Hb$ is a matrix valued function given by 
\beq
H(\yb)=\dpdx{}{\Bb}  \left[ 
\sum_i \log p_i\left(y_i\right) + \log|\det{\Bb}| + \frac{1}{T} \log p(\Bb)
\right]. 
\eeq

As an example, consider a uniform \apriori law for $\Bb$. 
Then we obtain the classical ML estimate which satisfies 
\beq \label{grad2}
\sum_{t} \Hb \left( \yb(t) \right) =\zmat 
\hbox{~~with~~} 
\Hb \left( \yb \right) = \phib(\yb) \, \yb^t - \Ib, 
\eeq
where $\phib(\yb)=[\phi_{1}(y_{1}),\ldots,\phi_{n}(y_{n})]^t$ with 
\beq
\phi_i(z)= - \frac{p'_i(z)}{p_i(z)}. 
\eeq

One can add some extra constraints to this optimization. For example, 
we can optimize the MAP criterion subject to the constraint 
$\frac{1}{T} \sum_{t}  \yb(t) \, \yb^t(t) = \Ib$ which leads again to 
\beq
\sum_{t} \Hb \left( \yb(t) \right) =\zmat 
\hbox{~with~} 
\Hb \left( \yb \right) = \alpha (\yb \, \yb^t-\Ib) 
+ \beta \left( \phib(\yb) \, \yb^t + \yb \, \phib^t(\yb) \right). 
\eeq

Note that in all these relations, $\phi_i(z)$ is related to the 
probability distribution of the source number $i$. The following 
table gives the expression of this function for a few known cases. 

\[
\barr{||l|l|l||} \hline\hline   
\hbox{Gauss}  
& \disp{ p(z)\propto \expf{-\alpha z^{2}} }  
& \phi(z)=2\alpha z  
\\[6pt] \hline 
\hbox{Laplace} & 
\disp{ p(z)\propto \expf{-\alpha |z|} }        
& \phi(z)=\alpha\hbox{sign}(z) 
\\[6pt] \hline   
\hbox{Cauchy}  & 
\disp{ p(z)\propto \frac{1}{1+(z/\alpha)^{2}} }
& \phi(z)= \frac{2 z/\alpha^{2}}{1+(z/\alpha)^{2}}  
\\[6pt] \hline 
\hbox{Gamma}   & 
\disp{ p(z)\propto z^{\alpha}\expf{-\beta z} } 
& \phi(z)=-\alpha/ z + \beta 
\\[6pt] \hline 
\hbox{sub-Gaussian law}   & 
\disp{ p(z)\propto \expf{-\frac{1}{2}z^2}\hbox{sech}^2(z) }
& \phi(z)=z + \tanh(z) 
\\[6pt] \hline 
\hbox{Mixture of Gaussians}   & 
\barr{@{}ll}
p(z) \propto & ~~\expf{-\frac{1}{2}(z-\alpha)^2}\\ 
             & +\expf{-\frac{1}{2}(z+\alpha)^2}
\earr  
& \phi(z)=\alpha z - \alpha\tanh(\alpha z) 
\\[6pt] \hline\hline 
\earr
\]

\bigskip\noindent{\bf Remark:} \\ 
$H(\yb)$ in equations (\ref{grad1}) and (\ref{grad2}) corresponds 
to the gradient of MAP and ML criteria. A common technique to obtain 
the MAP or the ML solutions is then to use a gradient based algorithm 
such as 
\beq
\Bb^{(k+1)} = \Bb^{(k)} - \gamma \Hb(\yb) 
\eeq
where $(k)$ and $(k+1)$ stand for two successive iterations in static 
case or two successive time instants for dynamic case. 
This equation forms the main body of a great number of neural 
network (NN) based algorithms for source separation. 

\bigskip
\subsection{Accounting for errors}

Here we relax the inversibility of the matrix $\Ab$ and take also 
account of the errors on the data. 
As an example, we consider the case where the errors can be 
modelled by an additive term $\epsilonb(t)$:
\beq \label{model2}
\xb(t)=\Ab \, \sb(t) + \epsilonb(t), \quad t=1,\ldots,T.
\eeq
We assume also that we can assign a probability law $p(\epsilonb)$ to 
$\epsilonb$. In general, it is natural to assume that $\epsilonb(t)$ has 
independent components and is centered and white, \ie 
\beq
\log p(\epsilon(1),\ldots,\epsilon(T)) 
= \sum_{t}\sum_i \log p_i\left( \epsilon_i(t) \right).
\eeq
From this assumption, we obtain 
\beq
\log \pxas  
= \sum_{t}\sum_i q_i\left( x_i(t) - [\Ab\sb]_i(t) \right)
\eeq
with $q_i(.)=\log p_i(.)$. 

\rem{
Again here we also assume that we can assign a probability law $p(\Ab)$ 
to the \mm $\Ab$ to translate any prior knowledge we have about 
(or we wish to impose to) it.
}
  
Now, we can give the expression of the posterior law which is 
\beqn
\log \pasx &=& \log \pxas + \log p(\sb_{1..T}) + \log p(\Ab) + cte 
\nonumber \\ 
&=& \sum_t \sum_i q_i\left( x_i(t) - [\Ab\sb]_i(t) \right) 
    + \log p(\sb_{1..T}) + \log p(\Ab) + cte. \nonumber \\ &  
\eeqn

As mentioned before, from here, we can go in at least three directions: 
\bit
\item First integrate $\pasx$ with respect to $\Ab$ to obtain 
$p(\sb_{1..T} | \xb_{1..T})$ and estimate $\sb_{1..T}$ by 
\beq
\sbh_{1..T} = \argmax{\sb_{1..T}}{p(\sb_{1..T} | \xb_{1..T})}. 
\eeq 

\item Second integrate $\pasx$ with respect to $\sb_{1..T}$ to obtain 
$p(\Ab | \xb_{1..T})$ and estimate $\Ab$ by 
\beq
\Abh = \argmax{\Ab}{p(\Ab | \xb_{1..T})}. 
\eeq 
But, here, when $\Abh$ is obtained, we still have to obtain 
$\Bbh=\Abh^{-1}$ 
and $\Abh$ may not be invertible. 

\item Third, optimize $\pasx$ simultaneously with respect to both 
$\sb_{1..T}$ and $\Ab$ by using an alternating optimization 
procedure such as 

\beq \label{alg1}
\left\{\barr{lcl}
\disp{ \sbh_{1..T}^{(k)} } 
&=& \disp{
\argmax{\sb_{1..T}}{ p\left( \Abh^{(k-1)},\sb_{1..T} | \xb_{1..T} \right) }
}
\\ 
\disp{ \Abh^{(k)} }
&=& \disp{
\argmax{\Ab}{ p\left( \Ab,\sbh_{1..T}^{(k-1)} | \xb_{1..T} \right) }
}
\earr\right.
\eeq
\eit

In the two first cases, the integrations can be done analytically 
only in the Gaussian case. 
We then obtain closed form expressions for the solutions. 

In any case, before applying any optimization, we have to 
ensure that the criterion to be optimized has at least an optimum 
and that this optimum is unique. 

\bigskip
\subsection{Spatially independent and white sources}

The case where we can assume that the sources are independent 
and white is the simplest one. We have:  
\[
\log p(\sb_{1..T})= \sum_t \sum_j r_j(s_j(t))
\] 
and 
\beq
\log \pasx =
  \sum_t \sum_i q_i\left( x_i(t) - z_i(t) \right) 
 + \sum_t \sum_j r_j(s_j(t)) 
 + \ln p(\Ab)  + cte. 
\eeq
with $z_i=[\Ab\sb]_i$. 
Then, we can omit the time summation. To simplify the details 
of the derivations, let first assume  
\beq
p(\Ab) \propto \expf{ -\frac{1}{2\sigma_a^2} \| \Ab \|^{2} } 
=\expf{ -\frac{1}{2\sigma_a^2} \sum_k\sum_l a_{k,l}^{2} }.
\eeq
Later, we will also consider other possibilities such as (\ref{pa2}), 
(\ref{pa3}) or (\ref{pa4}). 

\bigskip
\subsubsection{Joint MAP estimation}

\bigskip 
First we consider the joint estimation of $\Ab$ and $\sb$ where 
the alternating optimization algorithm becomes

\beq \label{alg2}
\left\{\barr{lcl}
\disp{ \sbh^{(k)} } 
&=& \disp{ \argmax{\sb}{ 
  \sum_i q_i\left( x_i - z_i \right) + \sum_j r_j(s_j)  
  } }
\\ 
\disp{ \Abh^{(k)} }
&=& \disp{ \argmax{\Ab}{ 
  \sum_i q_i\left( x_i - z_i \right) 
- \frac{1}{2\sigma_a^2} \sum_k \sum_l a_{kl}^2
  } }
\earr\right. 
\eeq 
The solution at each iteration has to satisfy 
\beq
\left\{\barr{l}
\disp{ \dpdx{}{s_j}
	   = -\sum_i a_{ij} \, q'_i\left( x_i - z_i \right) + r'_j(s_j) = 0
     }
\\ 
\disp{ \dpdx{}{a_{ij}}
      =-s_j \, q'_i\left( x_i - z_i \right) 
       - \frac{1}{\sigma_a^2} a_{ij} = 0
     }
\earr\right. 
\eeq 
These equations are in general nonlinear and depend on the expressions 
of $q$ and $r$. One exception is the 
particular case of Gaussian laws 
\[
p_i(n)\sim \Nc(0,\sigma_{\epsilon}^2) \lra q_i(n)
=-\frac{1}{2\sigma_{\epsilon}^2} n^2 
\lra q'_i(n)=-\frac{1}{\sigma_{\epsilon}^2} n
\] 
and
\[
p_j(s)\sim \Nc(0,\sigma_s^2) \lra r_j(s)=-\frac{1}{2\sigma_s^2} s^2 
\lra r'_i(s)=-\frac{1}{\sigma_s^2} s
\]
where we obtain two sets of linear equations to solve for 
$s_j$ and $a_{ij}$ : 

\beq 
\left\{\barr{l} 
\disp{\frac{1}{\sigma_{\epsilon}^2} 
      \sum_i a_{ij} \left(x_i-[\Ab\sb]_i\right)  
      - \frac{1}{\sigma_s^2} s_j = 0
     }
\\ 
\disp{\frac{1}{\sigma_{\epsilon}^2} 
      \, s_j \left(x_i - [\Ab\sb]_i\right)  
      - \frac{1}{\sigma_a^2} a_{ij} = 0
     }
\earr\right. 
\lra 
\left\{\barr{l} 
\disp{\sum_i a_{ij} \left(x_i-[\Ab\sb]_i\right)-\lambda s_j = 0}
\\ 
\disp{s_j \left(x_i-[\Ab\sb]_i \right) - \mu a_{ij} = 0}
\earr\right. 
\eeq
with $\lambda=\sigma_{\epsilon}^2/\sigma_s^2$ and  
$\mu=\sigma_{\epsilon}^2/\sigma_a^2$. 

These two equations have to be solved in each 
iteration of alternating optimization procedure. 
Two strategies can be used : 
\bit
\item Solve these equations for each $s_{j}$ and then for each $a_{ij}$ 
at each iteration:

\beq \label{sol0}
\left\{\barr{lcl} 
\disp{s_j=\frac{\sum_i a_{ij} (x_i-\wh{x}_i)}{\lambda+\|\ab_{j*}\|^2}
     } 
\\ 
\disp{a_{ij}
=\frac{s_j (x_i-\wh{x}_i)}{s_j^2+\mu} 
     }
\earr\right. 
\eeq
with $\wh{x}_i = \sum_{k\not= j} a_{ik} s_k$, and 
$\|\ab_{i*}\|^2=\sum_j a_{i,j}^2$.  
This is a single coordinate-wise gradient descent based algorithm. 

\item Solve these equations for all $s_{j}$ and then for all $a_{ij}$ 
at each iteration:

\beq \label{sol1}
\left\{\barr{lcl}
\Ab^t (\xb - \Ab \sb) - \lambda \sb &=& \bm{0} \\ 
(\xb - \Ab\sb) \sb^t -\mu \Ab       &=&  \bm{0}
\earr\right.
\lra 
\left\{\barr{lcl}
\sb &=& (\Ab^t\Ab+\lambda\Ib)^{-1} \Ab^t \xb \\   
\Ab &=& \xb \sb^t (\sb\sb^t+\mu\Ib)^{-1} \\ 
    &=&\frac{\xb\,\sb^t}{\mu}
	    \left[\Ib -\frac{\sb\,\sb^t}{\sb^t\sb+\mu}\right] 
\earr\right. 
\eeq
This is a bloc coordinate-wise gradient descent based algorithm. 

\eit

\bigskip\noindent{\bf Remark 1:}\\ 
These two last closed form expressions give us the possibility to discuss 
the convergency of the joint MAP algorithm in the considered Gaussian case. 
We may immediately note that $\Ab$ obtained by this algorithm is not 
invertible. This means that, in the Gaussian hypothesis, this algorithm 
does not really separate the signals. Actually, we could remark this from 
the expression of the joint criterion in this case which is 
\beq
J(\Ab,\sb)=-\log p(\Ab,\sb | \xb) 
= \| \xb -\Ab\sb \|^2 + \lambda \| \sb \|^2 + \mu \| \Ab \|^2 + cte.
\eeq
As we can see, in this case, $J(\Ab,\sb)$ is a quadratic function of $\sb$ 
for given $\Ab$ and a quadratic function of $\Ab$ for given $\sb$, but 
it is a biquadratic function of both $\Ab$ and $\sb$. 
This symetry property means 
that the joint MAP solution in this case is not unique. 
This criterion may even have an infinite equivalent optima. 
The proposed iterative algorithm may then converge to any of these solutions 
depending on the initialization. 

Unfortunately, with any non Gaussian hypothesis, we can not obtain 
any closed form solution and the existance and the uniqueness of a 
global optimum is very hard to study. However, we can always propose 
either a fixed point or a gradient descent based algorithm to compute 
them numerically. 
For example, if we assume a Gaussian law for the noise, but non Gaussian 
prior laws for $\sb$ and for $\Ab$, we have
\beq
J(\Ab,\sb)=-\log p(\Ab,\sb | \xb) 
= \| \xb -\Ab\sb \|^2 + \lambda \phi(\sb) + \mu \psi(\| \Ab \|) + cte.
\eeq
where $\phi(\sb)\propto -\log p(\sb)$ and $\psi(\sb)\propto -\log p(\Ab)$. 
Note that the choice of these prior laws is then important if we want to 
eliminate the above mentionned symetry property and to be able to find a 
unique solution to the problem. Then, a gradient based algorithm writes:
\beq
\left\{\barr{lcl}
\sbh^{(k+1)}&=&\sbh^{(k)} - \alpha^{(k)} \,\,  
\dpdx{J}{\sb}(\Abh^{(k)},\sbh^{(k)}) \\ 
\Abh^{(k+1)}&=&\Abh^{(k)} - \beta^{(k)} \,\,  
\dpdx{J}{\Ab}(\Abh^{(k)},\sbh^{(k)})
\earr\right.
\eeq
where $\alpha^{(k)}$ and $\beta^{(k)}$ are two step parameters which can 
be either constant (fixed step gradient) or adaptive during the iterations 
$(k)$. Replacing for the gradient expressions we obtain:
\beq
\left\{\barr{lcl}
\sbh^{(k+1)}&=& \sbh^{(k)}+ \alpha^{(k)} 
\left[
2 \Abh^{t^{(k)}} (\xb-\xbh^{(k)}) + \lambda \dpdx{\phi}{\sb}(\sbh^{(k)})
\right]
\\ 
\Abh^{(k+1)}&=& \Abh^{(k)}+ \beta^{(k)} 
\left[
2 \sbh^{(k)}(\xb-\xbh^{(k)})^t + \mu \dpdx{\psi}{\Ab}(\Abh^{(k)})
\right]
\earr\right.
\eeq
with $\xbh^{(k)}=\Abh^{(k)}\sbh^{(k)}$. 

A fixed point based algorithm writes:
\beq
\left\{\barr{lcl}
\dpdx{\phi}{\sb}(\sbh^{(k)})&=& \frac{-1}{\lambda} 
\left[
\Abh^{t^{(k)}} (\xb-\xbh^{(k)})
\right]
\\ 
\dpdx{\psi}{\Ab}(\Abh^{(k)})&=& \frac{-1}{\mu}  
\left[
\sbh^{(k)}(\xb-\xbh^{(k)})^t
\right]
\earr\right.
\eeq
One can make the comparison with different neural network based algorithm. 

\bigskip\noindent{\bf Remark 2:}\\ 
In the Gaussian hypothesis case, if we use the prior law (\ref{pa2}), we 
obtain similar closed form expressions equivalent to (\ref{sol0}) 
\beq \label{sol02}
\left\{\barr{lcl} 
\disp{s_j=\frac{\sum_i a_{ij} (x_i-\wh{x}_i)}{\lambda+\|\ab_{j*}\|^2}
     } 
\\ 
\disp{
a_{ij}=\frac{s_j (x_i-\wh{x}_i)}{s_j^2+\mu},\, \hbox{for~} i= j 
\quad\hbox{and}\quad 
a_{ij}=\frac{s_j (x_i-\wh{x}_i)}{s_j^2-\mu}, \, \hbox{for~} i\not= j. 
     }
\earr\right. 
\eeq

\bigskip\noindent{\bf Remark 3:}\\ 
The main interest of this approach is that we can, at least in theory,  
account for the existance of any correlation between $s_j$ and $s_k$ 
or to model the temporal behavior of any source $s_j(t)$, for example, 
via a markov model. We can also account for any prior information 
we may have about the mixing matrix $\Ab$ or impose any desired 
structure for the separating matrix $\Bb$. 
For example, if we know that the sources are labelled in such a way 
that the sensor $x_i$ is closer to the sources  
$s_i$, $s_{i-1}$ and $s_{i+1}$ than to any others, we can use it 
by choosing a prior probability law  
\beq
p(\Ab) \propto \expf{-\frac{1}{\sigma_a^2}\sum_i\sum_j(w_{ij} a_{ij})^2}
\eeq
with 
\beq 
w_{ij}=\left\{\barr{lcl}
 1            & if & i=j \\ 
 1/(2|i-j+1]) & if & i\not= j 
\earr\right.
\eeq
or $w_{ii}=1$, $w_{i,i-1}=w_{i-1,i}=\alpha$ and $(1-\alpha)$ 
for all the other coefficients $w_{ij}$ for some $0.5 < \alpha < 1$. 
Then the equations (\ref{sol0}) and (\ref{sol1}) become 
\beq \label{sol3}
\left\{\barr{lcl} 
\disp{s_j=\frac{\sum_i a_{ij} (x_i-\wh{x}_i)}{\lambda+\|\ab_{i*}\|^2}} 
\\ 
\disp{a_{ij}=\frac{s_j (x_i-\wh{x}_i)}{w_{ij}^2(s_j^2+\mu)}}
\earr\right. 
\eeq
and
\beq \label{sol4}
\left\{\barr{lcl}
\sb &=& (\Ab^t\Ab+\lambda\Ib)^{-1} \Ab^t \xb \\   
\Ab &=& \xb \sb^t (\sb\sb^t+\mu \Wb\Wb^t)^{-1} 
\earr\right. 
\eeq

\bigskip
\subsubsection{Marginal MAP estimations}

\bigskip
Now, we consider the two other approaches of marginal MAP estimations.
First we note that we can 
rewrite $\xb=\Ab\sb$ with $\Ab$ is a $(m \times n)$ matrix as  
\beq
\xb=\Ab\sb=\Sb\ab
\eeq
where $\Sb$ a $(m \times m n)$ bloc Toeplitz matrix 
and $\ab$ a vector of 
dimension $m n$ obtained by pilling up all the rows of the matrix $\Ab$:
\beq
\Sb=\pmatrix{
\sb^t  & \zerob & \cdots &       & \zerob \cr
\zerob & \sb^t  & \cdots &       & \vdots \cr
\vdots &        &                &        \cr
\zerob & \cdots &        & \sb^t &\zerob \cr
\zerob & \zerob & \cdots &       & \sb^t }
\quad\hbox{and}\quad 
\ab=\pmatrix{\ab_{1*} & \ab_{2*} & \cdots & \ab_{m*}}^t
\eeq

To be able to obtain closed form expression, in the following 
we consider only the Gaussian case: 
\beqn
p(\xb | \Ab,\sb) &\propto& 
\expf{-\frac{1}{2\sigma_{\epsilon}^2} \| \xb -\Ab\sb \|^2} \\ 
p(\Ab) &\propto& 
\expf{-\frac{1}{2\sigma_{a}^2} \psi(\Ab)} 
\hbox{~~with~~} \psi(\Ab)=\| \Ab \|^2 \\ 
p(\sb) &\propto& 
\expf{-\frac{1}{2\sigma_{s}^2} \phi(\sb)}
\hbox{~~with~~} \phi(\sb)=\| \sb \|^2
\eeqn
which gives 
\beq 
p(\Ab,\sb | \xb) \propto 
\expf{-\frac{1}{2\sigma_{\epsilon}^2} J(\Ab,\sb)} 
\eeq
with
\beqn
J(\Ab,\sb) 
&=& 
\| \xb -\Ab\sb \|^2 + \lambda \phi(\sb) + \mu \psi(\Ab) = 
\| \xb -\Ab\sb \|^2 + \lambda \| \sb \|^2 + \mu \| \Ab \|^2~\quad~~
\\ 
&=& \| \xb -\Sb\ab \|^2 + \lambda \phi(\Sb) + \mu \psi(\ab) =
\| \xb -\Sb\ab \|^2 + \frac{\lambda}{m} \| \Sb \|^2 + \mu \| \ab \|^2~\quad~~
\eeqn
Note that, in this case, $J(\Ab,\sb)$ is a quadratic function of $\sb$ 
for fixed $\Ab$ and a quadratic function of $\Ab$ for fixed $\sb$, but 
it is a bilinear function of both $\Ab$ and $\sb$. This remark means 
that the joint MAP solution in this case is not unique. 
 
Note also that $J(\Ab,\sb)$ can be rewritten as
\beq
J(\Ab,\sb) 
= (\sb -\sbh)^t \wh{\Pb}_{s}^{-1} (\sb -\sbh) 
    - \sbh^t\wh{\Pb}_{s}^{-1}\sbh + \xb^t\xb + \mu \| \Ab \|^2  
\eeq
with
\[
\sbh=(\Ab^t\Ab+\lambda\Ib)^{-1}\Ab^t\xb
\quad \hbox{and}\quad  
\wh{\Pb}_{s}=(\Ab^t\Ab+\lambda\Ib)^{-1}; 
\]
or as
\beq
J(\ab,\Sb) = (\ab -\wh{\ab})^t \wh{\Pb}_{a}^{-1} (\ab -\wh{\ab}) 
    - \abh^t\wh{\Pb}_{a}^{-1}\abh + \xb^t\xb + \lambda \|\Sb\|^2 
\eeq
with 
\[
\wh{\ab}=(\Sb^t\Sb+\mu\Ib)^{-1}\Sb^t\xb 
\quad \hbox{and}\quad  
\wh{\Pb}_{a}=(\Sb^t\Sb+\mu\Ib)^{-1}. 
\]  
With these notations, it is then easy to obtain the expression of the 
marginals laws $p(\sb|\xb)$ and $p(\Ab|\xb)$: 
\beqn
-\ln p(\Ab|\xb) &\propto& 
  -\ln \left| \det{\wh{\Pb}_s^{-1}}\right| - J(\Ab,\sbh) \nonumber\\ 
&=& -\ln \left| \det{\Ab^t\Ab+\lambda\Ib}\right| 
    - \xb^t (\xb - \Ab\sbh) + \mu \| \Ab \|^2   
\\ 
-\ln p(\sb|\xb) &\propto& 
-\ln \left| \det{\wh{\Pb}_a^{-1}}\right| - J(\abh,\sb) \nonumber\\ 
&=& -\ln \left| \det{\Sb^t\Sb+\mu\Ib}\right|
    -\xb^t (\xb -\Sb\abh) + \lambda \sb^t\sb 
\eeqn
In non Gaussian hypothesis for $\sb$ and $\ab$ where 
$-\ln p(\sb)=\phib(\sb)+cte$ and $-\ln p(\ab)=\psib(\ab)+cte$, 
we can always use the Laplace approximation of the posterior laws. 
Then, we can use the first lines of these two last equations by 
replacing $\det{\wh{\Pb}_s^{-1}}$ and $\det{\wh{\Pb}_a^{-1}}$ 
by, the jacobians of their respective log probability densities: 
$\nabla^2_{\sb} J(\Ab,\sb)$ 
and 
$\nabla^2_{\ab} J(\ab,\Sb)$ 
computed for the MAP estimates 
$\sbh=\argmin{\sb}{J(\Ab,\sb)}$ and $\abh=\argmin{\ab}{J(\ab,\Sb)}$. 

The marginal MAP solutions of $\Ab$ and $\sb$ respectively 
satisfy $\dpdx{\ln p(\Ab|\xb)}{\Ab} = 0$ and $\dpdx{\ln p(\sb|\xb)}{\sb} = 0$. 
Note that, excepted the Gaussian case where these equations have 
analytical solutions, we need a numerical optimisation algorithm to compute 
the solutions. 
For example, the marginal MAP estimate of $\Ab$ can be computed by the 
following iterative algorithm: 
\beq \label{MMAP}
\barr{lcl}
\disp{ \Abh^{(k)} }
&=& \disp{ \argmax{\Ab}{ \ln \left| \det{\wh{\Pb}_s^{(k-1)}} \right| 
+J(\Ab,\sbh^{(k-1)})}} \\ 
&=&  \disp{ \argmax{\Ab}{ \ln \left| \det{\Ab^t\Ab+\lambda\Ib} \right| 
+\xb^t\Ab\sbh^{(k-1)} + \mu \psi(\Ab)}
  }
\earr
\eeq 
where
\beq
\sbh^{(k)} 
 = \argmin{\sb}{J(\Abh^{(k-1)},\sb)} 
 = \argmin{\sb}{\| \xb -\Abh^{(k-1)}\sb \|^2 + \lambda \phi(\sb)}. 
\eeq 

\rem{Note also that the difference between the joint MAP criterion  
and the marginal MAP criterion is 
$\det{\wh{\Pb}_s^{-1}}=\nabla^2_{\sb} J(\Ab,\sb)$. 
}
In general, neither of these equations have explicite solutions and we have to do the optimization numerically. For example, a gradient based algorithm to compute the marginal MAP estimate of $\Ab$ writes: 
\beq \label{MMAPG}
\left\{\barr{lcl}
\disp{ \Delta\sbh^{(k)} } 
&\propto& \disp{ \Abh^{t^{(k-1)}}(\xb-\Abh^{(k-1)}\sb)+\lambda \phib'(\sb) }
\\ 
\disp{ \Delta\Abh^{(k)} }
&\propto& \disp{ \Ab^t(\Ab^t\Ab+\lambda\Ib)^{-1}  
+\xb\sbh^{(k-1)} + \mu \psib'(\Ab)}
\earr\right. 
\eeq 

\bigskip
\subsection{Spatially correlated sources}

As a first extension we still assume that the sources are white, 
but they are spatially correlated, \ie 
\[
\log p(\sb_{1..T})= \sum_t r(s_1(t),\ldots,s_N(t))
\] 
where $r(s_1,\ldots,s_N)$ represents the joint probability law of 
the sources. Then we obtain
\beq
\log \pasx =
  \sum_t \sum_i 
  \left[q_i\left( x_i(t) - y_i(t) \right) +  r(s_1(t),\ldots,s_N(t))\right] 
+ \log p(\Ab) + cte. 
\eeq
But the main difficulty here will be the modeling of these 
dependencies and simplification of the expression of the 
joint probability law $r(s_1,\ldots,s_N)$. 
For example, if the sources are labelled in such a 
way that only the immediate neighbor sources are correlated, 
then we can use a first order Markov model and write 
\beq
r(s_1,\ldots,s_N)=\sum_j r(s_j | s_{j-1}, s_{j+1})
\eeq
However, the complexity of the optimization algorithms depend on the
expression of the potential function $r(s_j | s_{j-1}, s_{j+1})$. 
A simple case is a Gaussian model where 
\[
\sum_j r(s_j | s_{j-1}, s_{j+1}) 
= \frac{1}{2\sigma_s^2} \left| 2 s_j -(s_{j-1}+s_{j+1}) \right|^2 
= \frac{1}{2\sigma_s^2} \| \Db \sb \|^2
\]
with $\Db$ a tri-diagonal Toeplitz matrix with diagonal elements equal to 2 
and off-diagonal elements equal to -1. In this case, the equations 
(\ref{sol1}) become 
\beq \label{sol2}
\left\{\barr{lcl}
\Ab^t (\xb - \Ab \sb) - \lambda \Db^t\Db \sb 
&=& \bm{0} \\ 
(\xb - \Ab\sb) \sb^t - \mu \Ab 
&=& \bm{0}
\earr\right.
\lra 
\left\{\barr{lcl}
\sb &=& (\Ab^t\Ab+\lambda\Db^t\Db)^{-1} \Ab^t \xb \\   
\Ab &=& \xb \sb^t (\sb\sb^t + \mu \Ib)^{-1} 
\earr\right. 
\eeq

\bigskip
\subsection{Spatially independent but colored sources}

Here we assume that the sources are mutually independent but that they 
are temporally colored, \ie 
\[
\log p(\sb_{1..T})= \sum_j r_j(s_j(1),\ldots,s_j(T))
\] 
where $r_j(s_j(1),\ldots,s_j(T))$ represents the joint probability 
law of the different samples of source number $j$.  
Then we obtain
\beqn
\log \pasx &=& 
  \sum_t \sum_i q_i\left( x_i(t) - y_i(t) \right) 
  + \sum_j r_j(s_1(1),\ldots,s_j(T)) \nonumber \\ 
&& + \sum_k \sum_l a_{kl}^2 + cte. 
\eeqn
Here again, the main difficulty is the modelization  
and simplification of the expression of the joint probability laws  
$r_j(s_j(1),\ldots,s_j(T))$. 
For example, we can use a first order markov chain model and write 
\beq
r_j(s_j(1),\ldots,s_j(T))=\sum_t r_j(s_j(t) | s_j(t-1))
\eeq
As an example here we consider the Gaussian case 
(or equivalently first order AR models) :

\beq
r_j(s_j(1),\ldots,s_j(T))=-\sum_t \alpha_j (s_j(t) - s_j(t-1))^2
\eeq

With this assumption we have 
\beq \label{alg3}
\left\{\barr{lcl}
\disp{ \sbh_{1..T}^{(k)} } 
&=& \disp{ \argmax{\sb}{ 
  \sum_t \sum_i q_i\left( x_i(t) - y_i(t) \right) 
- \sum_t \sum_j \alpha_j (s_j(t)- s_j(t-1))^2  
  } } 
\\ 
\disp{ \Abh^{(k)} }
&=& \disp{ \argmax{\Ab}{ 
  \sum_t \sum_i q_i\left( x_i(t) - y_i(t) \right) 
+ \frac{1}{\sigma_{a}^{2}} \sum_k \sum_l a_{kl}^2
  } }
\earr\right. 
\eeq 
The solution at each iteration has to satisfy 
\beq
\left\{\barr{lcl}
\disp{ \sum_t 
    \sum_i a_{ij} \, q'_i\left( x_i(t) - y_i(t) \right) 
  - \sum_t \sum_j 2 \alpha_j \, (s_j(t) - s_j(t-1)) 
    } = 0
\\ 
\disp{ \sum_t s_j \, q'_i\left( x_i(t) - y_i(t) \right) 
           + \frac{2}{\sigma_{a}^{2}} a_{ij}
    } = 0
\earr\right. 
\eeq 
For the special case of Gaussian noise we obtain  

\beq
\left\{\barr{lcl} 
\disp{  
\sum_t \sum_i a_{ij} \left( x_i(t) - [\Ab\sb]_i(t) \right)  
- \sum_t \sum_j 2 \lambda_j (s_j(t)- s_j(t-1)) } &=& 0
\\ 
\disp{  
\sum_t s_j(t) \left( x_i(t) - [\Ab\sb]_i(t) \right)  
+ \mu a_{ij} } &=& 0
\earr\right. 
\eeq
with $\lambda_j=\alpha_j\sigma_{\epsilon}^2$ and 
$\mu=\frac{2 \sigma_{\epsilon}^2}{\sigma_{a}^{2}}$. 

The two algorithms of (\ref{sol0}) and (\ref{sol1}) in this case 
become:

\beq 
\left\{\barr{lcl} 
\disp{s_j(t)=\frac{\lambda_{j} s_{j}(t-1)+\sum_i a_{ij} 
(x_i-\wh{x}_i)}{\lambda+\|\ab_{i*}\|^2}
     } 
\\ 
\disp{a_{ij}
=\frac{s_j (x_i-\wh{x}_i)}{s_j^2+\mu} 
     }
\earr\right. 
\eeq
and
\beq \label{alg4}
\left\{\barr{lcl}
\sb(t) &=& (\Ab^t\Ab+\lambda\Ib)^{-1} 
\left[\diag{\lambda_1,\ldots,\lambda_n} \sb(t-1) + \Ab^t \xb(t)\right] \\   
\Ab &=& \xb \sb^t (\sb\sb^t+\mu\Ib)^{-1} 
\earr\right. 
\eeq

Here, we conclude the presentation of the Bayesian approach to source separation. Beside some of the details of implementation, we showed that 
the Bayesian approach gives us the possibility to push further some of the 
limitations of the classical techniques in source separation. 
In the next section we give a few preliminary numerical results to show 
the performances of the proposed algorithems. 

\section{Simulation results}

In the following we give a few preliminary examples of simple source 
separation problem to show some performances of the proposed methods. 
In all these examples, we used the following algorithm:
\beq \label{UsedAlg}
\left\{\barr{lcl}
\yb &=& (\Ab^t\Ab+\lambda\Ib)^{-1} \Ab^t \xb \\ 
\sb &=& \gb(\yb)\\ 
\Delta\Ab 
&\propto& \Ab^t(\Ab^t\Ab+\lambda\Ib)^{-1}  
+\xb\sb + \mu \psib'(\Ab)
\earr\right. 
\eeq 
with $\lambda=\mu=.1$, $N=100$ and appropriate $\gb$.  

\bigskip
\subsection{Example 1}
Hier, we considered two sources 
\beq
\left\{ 
\barr{lcl} 
s_1(t) &=& \sin(500 t+10 \cos(50 t)) \\ 
s_2(t) &=& \sin(300 t)
\earr
\right., 
\quad t=[0:.001:.499].
\eeq
and used the mixing matrice 
$\Ab=\pmatrix{1 & .4 \cr -.6 & 1}$ 
to obtain the two set of data $x_1(t)$ and $x_2(t)$. 
Then we applied the algorithm given in (\ref{UsedAlg}). 
The following figures show the obtained results 
$\wh{s}_1(t)$ and $\wh{s}_2(t)$.

\bcc
\btabu[b]{@{}cc@{}}
\btabu[b]{@{}c@{}}
$\left\{\barr{@{}l@{}}
s_1(t) \\ ~\\ ~\\  
s_2(t)  
\earr\right.$  
~\\ ~\\ ~\\ 
$\left\{\barr{@{}l@{}}
x_1(t) \\ ~\\ ~\\  
x_2(t)  
\earr\right.$ 
~\\ ~\\ ~\\ 
$\left\{\barr{@{}l@{}}
\wh{s}_1(t) \\ ~\\ ~\\  
\wh{s}_2(t)     
\earr\right.$ 
\\ ~\\ 
\etabu
&
\btabu[b]{@{}c@{}}
\epsfxsize=100mm\epsfysize=75mm
\epsfbox{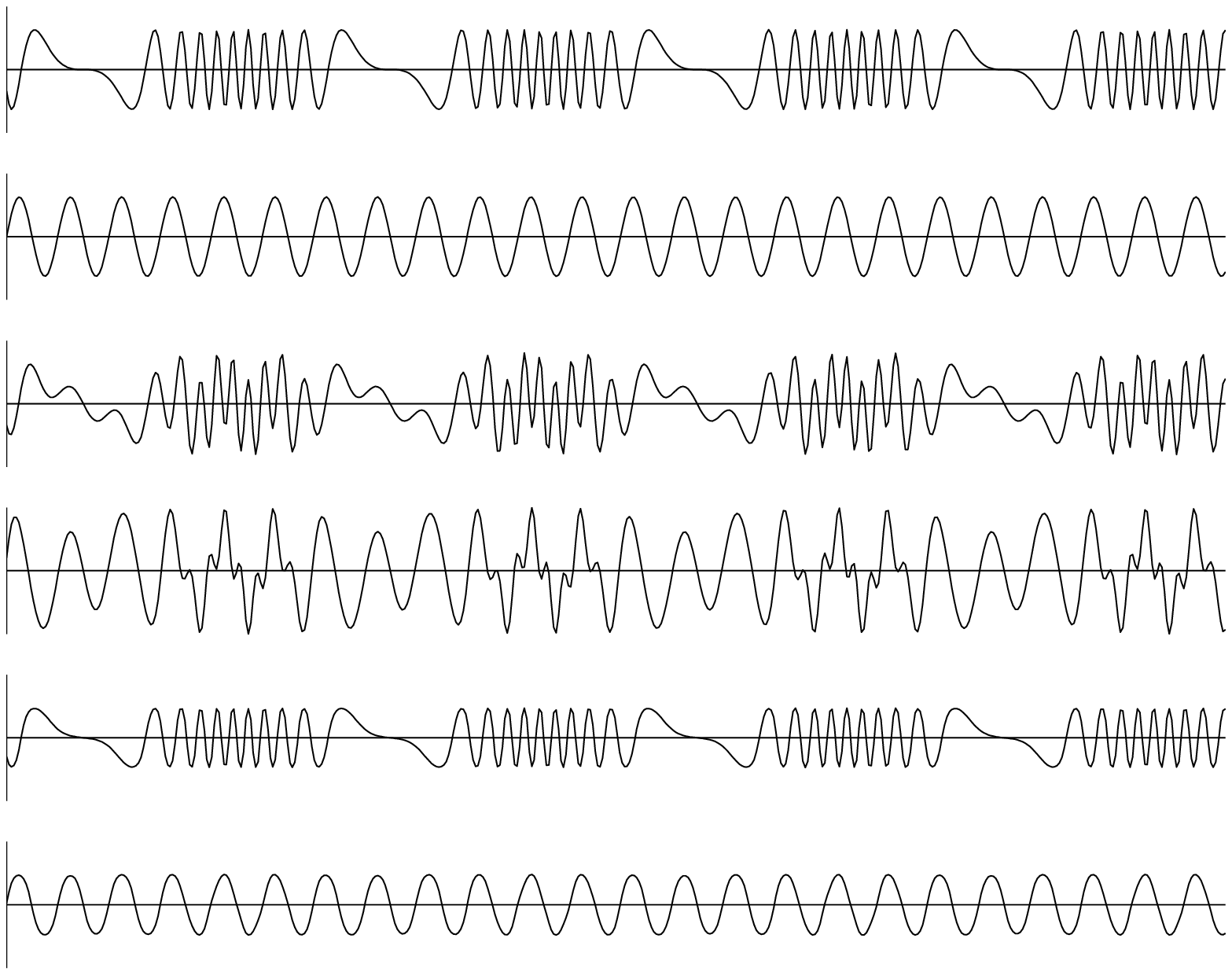} 
\etabu
\etabu
\\ {\small Fig.~1: Results of the source separation in Example 1.}
\ecc

\bcc
\btabu{@{}c@{}}
\epsfxsize=120mm\epsfysize=40mm
\epsfbox{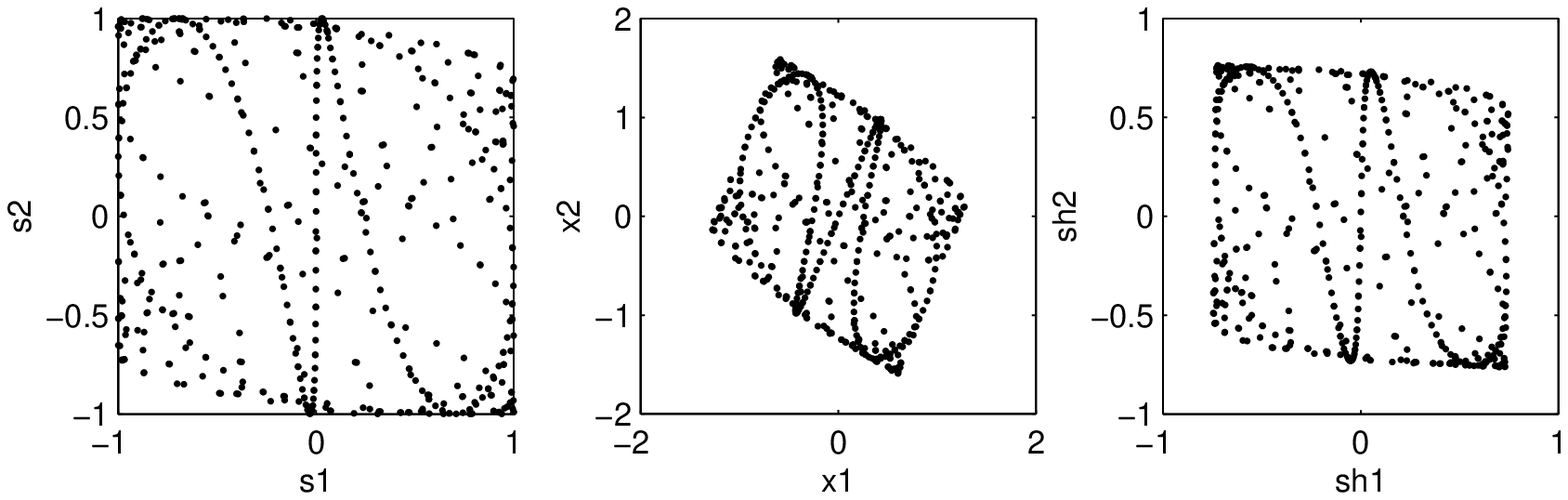} 
\etabu
\\ {\small Fig.~2: Results of the source separation in Example 1: 
phase space distribution of sources, mixed signals and separated sources}
\ecc

\bcc
\btabu{@{}c@{}}
\epsfxsize=120mm\epsfysize=60mm
\epsfbox{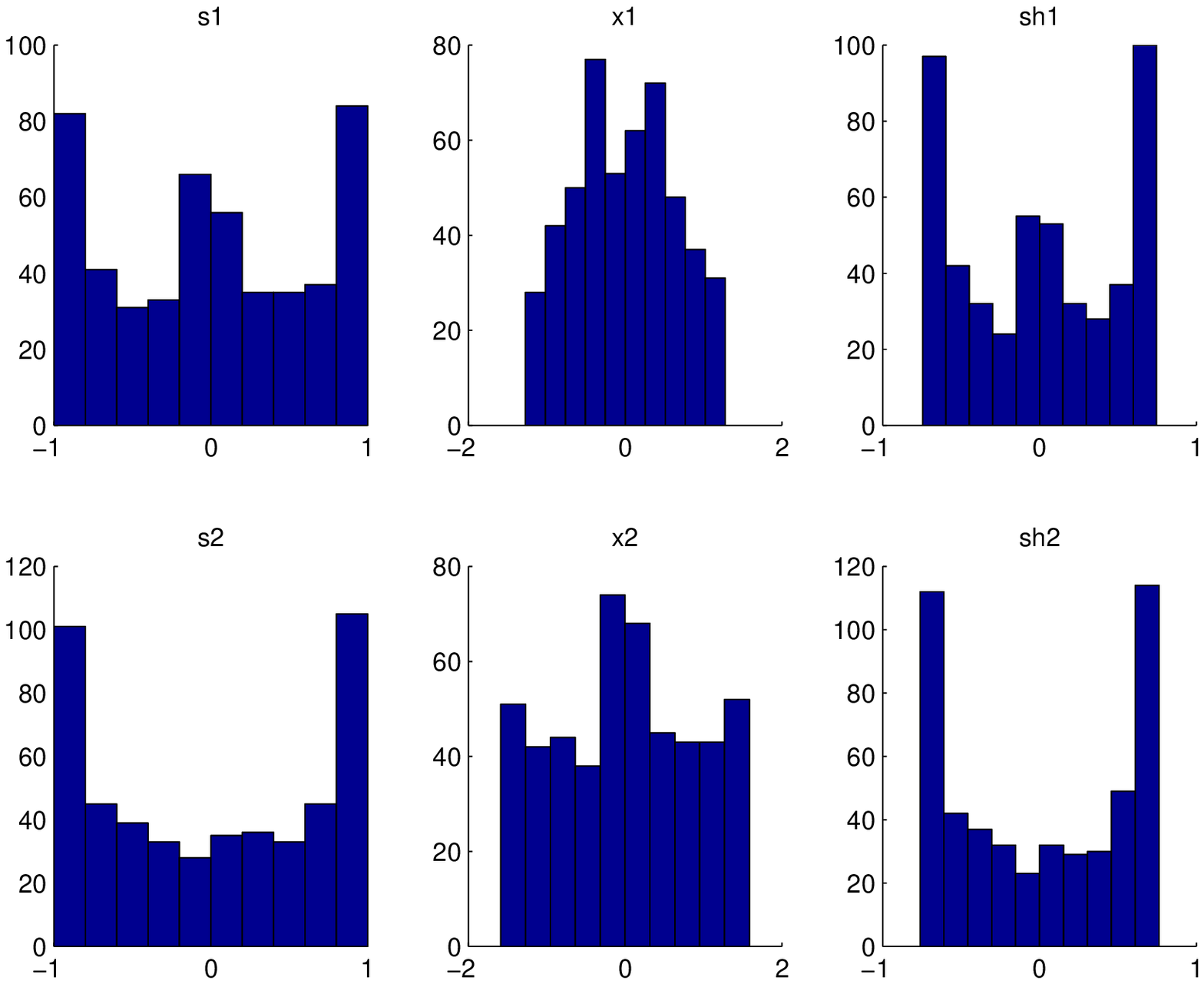} 
\etabu
\\ {\small Fig.~3: Results of the source separation in Example 1: 
histograms of sources, mixed signals and separated sources}
\ecc

The two sources are well separated. 

\bigskip
\subsection{Example 2}
Hier, we considered three sources 
\beq
\left\{ 
\barr{lcl} 
s_1(t) &=& \sin(500 t+10 \cos(50 t)) \\ 
s_2(t) &=& \sin(300 t) \\ 
s_3(t) &=& \sign(\cos(120 t-5 cos(50 t))) 
\earr
\right., 
\quad t=[0:.001:.499].
\eeq
and used the mixing matrice    
\[
\Ab=
.3*\pmatrix{
1. & -.5 & .2  \cr 
-.5 & 1. & -.5 \cr 
.5 & -.5 & 1. }
\]
to obtain the three set of data 
$x_1(t)$, $x_2(t)$ and $x_3(t)$. 
The following figures show the obtained results. 

\bcc
\btabu[b]{@{}cc@{}}
\btabu[b]{@{}c@{}}
$\left\{\barr{@{}l@{}}
s_1(t) \\ ~\\ ~\\  
s_2(t) \\ ~\\ ~\\  
s_3(t)
\earr\right.$  
\\ ~\\ ~\\  
$\left\{\barr{@{}l@{}}
x_1(t) \\ ~\\ ~\\  
x_2(t) \\ ~\\ ~\\  
x_3(t) 
\earr\right.$
\\ ~\\ ~\\  
$\left\{\barr{@{}l@{}}
\wh{s}_1(t) \\  ~\\ ~\\  
\wh{s}_2(t) \\  ~\\ ~\\  
\wh{s}_3(t) 
\earr\right.$
\\ ~\\ 
\etabu
&
\btabu[b]{@{}c@{}}
\epsfxsize=100mm\epsfysize=112.5mm
\epsfbox{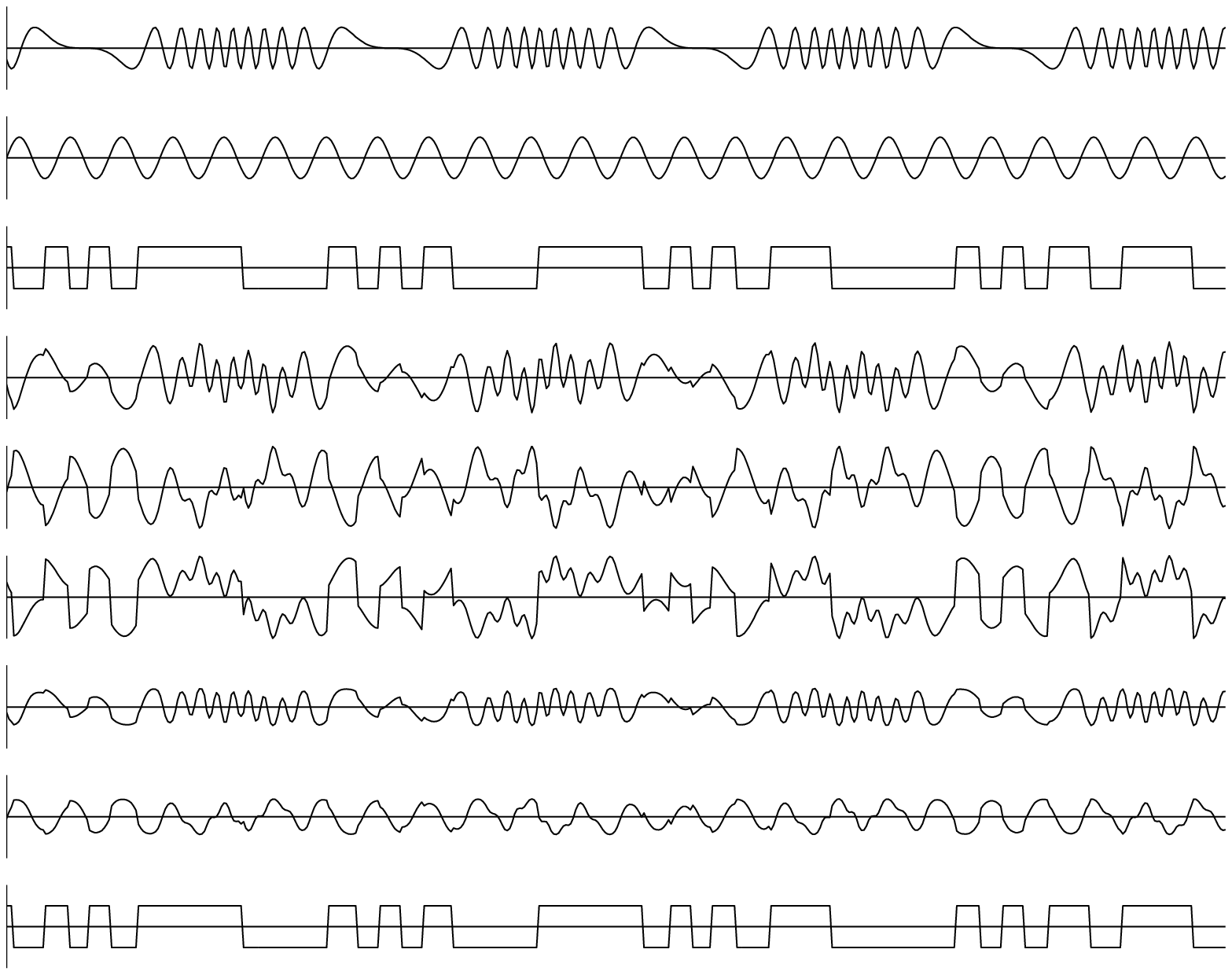} 
\etabu
\etabu
\\ {\small Fig.~4: Results of the source separation in Example 2.}
\ecc

\bcc
\btabu{@{}c@{}}
\epsfxsize=120mm\epsfysize=40mm
\epsfbox{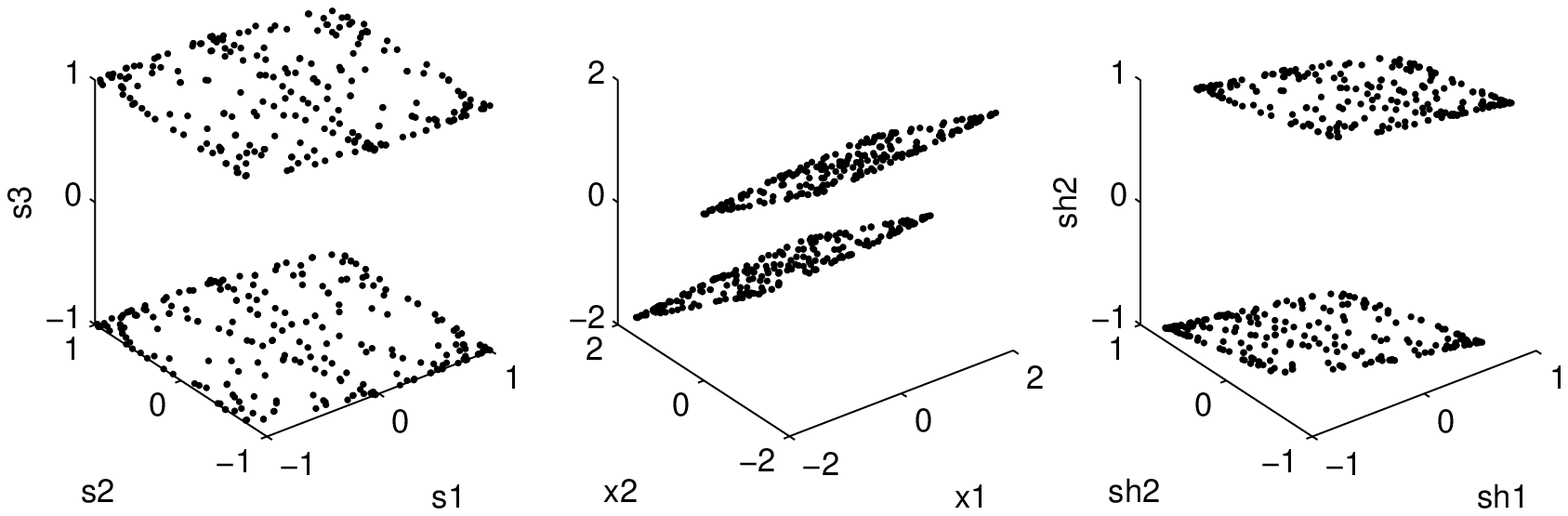} 
\etabu
\\ {\small Fig.~5: Results of the source separation in Example 2: 
phase space distribution of sources, mixed signals and separated sources}
\ecc

\bcc
\btabu{@{}c@{}}
\epsfxsize=120mm\epsfysize=90mm
\epsfbox{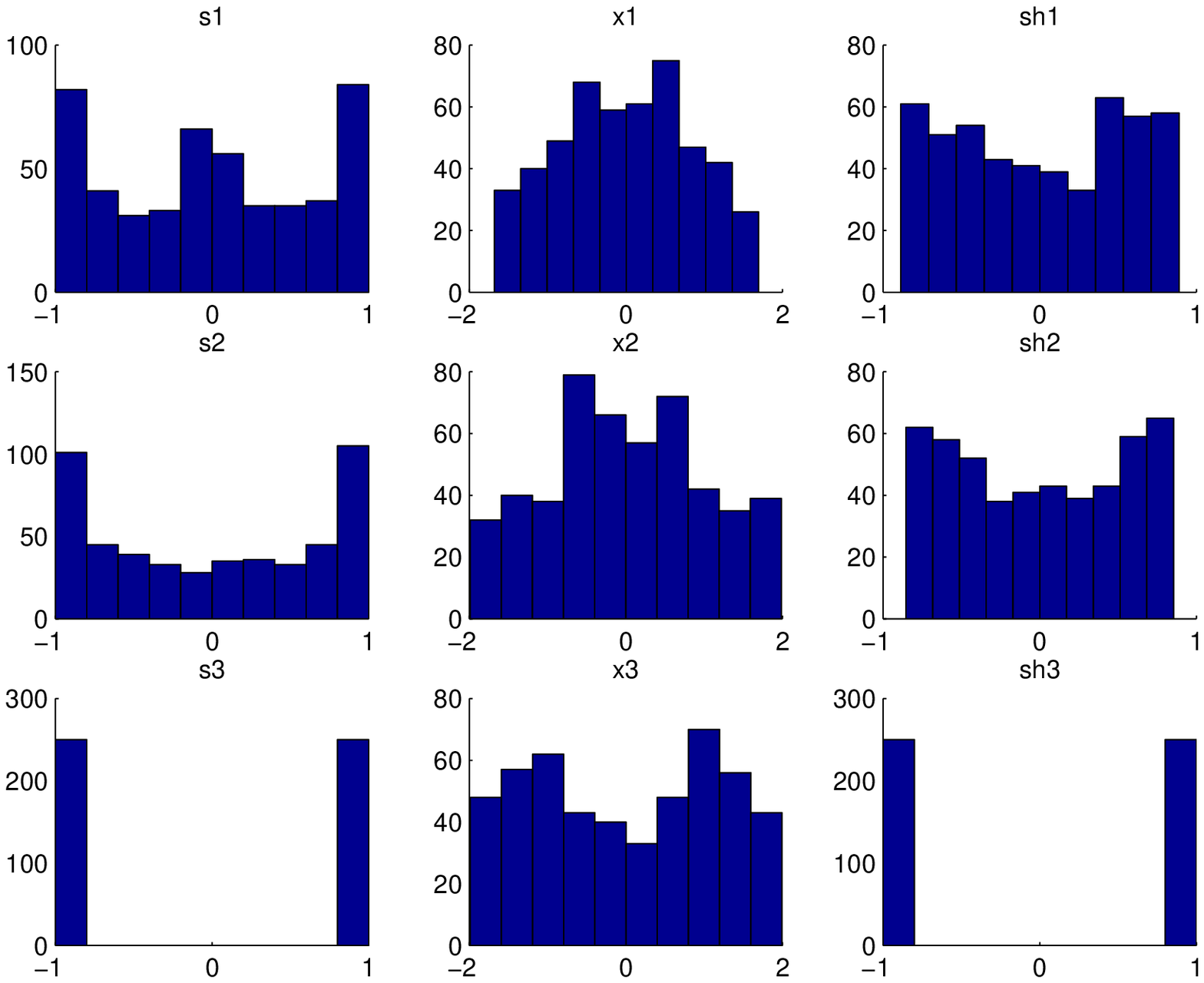} 
\etabu
\\ {\small Fig.~6: Results of the source separation in Example 2: 
histograms of sources, mixed signals and separated sources}
\ecc
Here also the three sources are well separated. 

\bigskip
\subsection{Example 3}
Hier, we considered the two sources of the first example, but we simulated the case where there are three receivers using the mixing matrice    
\[
\Ab=
\pmatrix{
1.  & -.5 \cr 
 .5 & 1. \cr
-.2 & .5 }
\]
to obtain the three set of data $x_1(t)$, $x_2(t)$ and $x_3(t)$. 
Then we applied again the algorithm given in (\ref{UsedAlg}). 
The following figures show the obtained results. 

\bcc
\btabu[b]{@{}cc@{}}
\btabu[b]{@{}c@{}}
$\left\{\barr{@{}l@{}}
s_1(t) \\ ~\\ ~\\  
s_2(t) 
\earr\right.$
\\ ~\\ ~\\ ~\\  
$\left\{\barr{@{}l@{}}
x_1(t) \\ ~\\ ~\\   
x_2(t) \\ ~\\ ~\\   
x_3(t) 
\earr\right.$
\\ ~\\ ~\\ ~\\ 
$\left\{\barr{@{}l@{}}
\wh{s}_1(t) \\  ~\\ ~\\  
\wh{s}_2(t)  
\earr\right.$
\\ ~\\  
\etabu
&
\btabu[b]{@{}c@{}}
\epsfxsize=100mm\epsfysize=100mm
\epsfbox{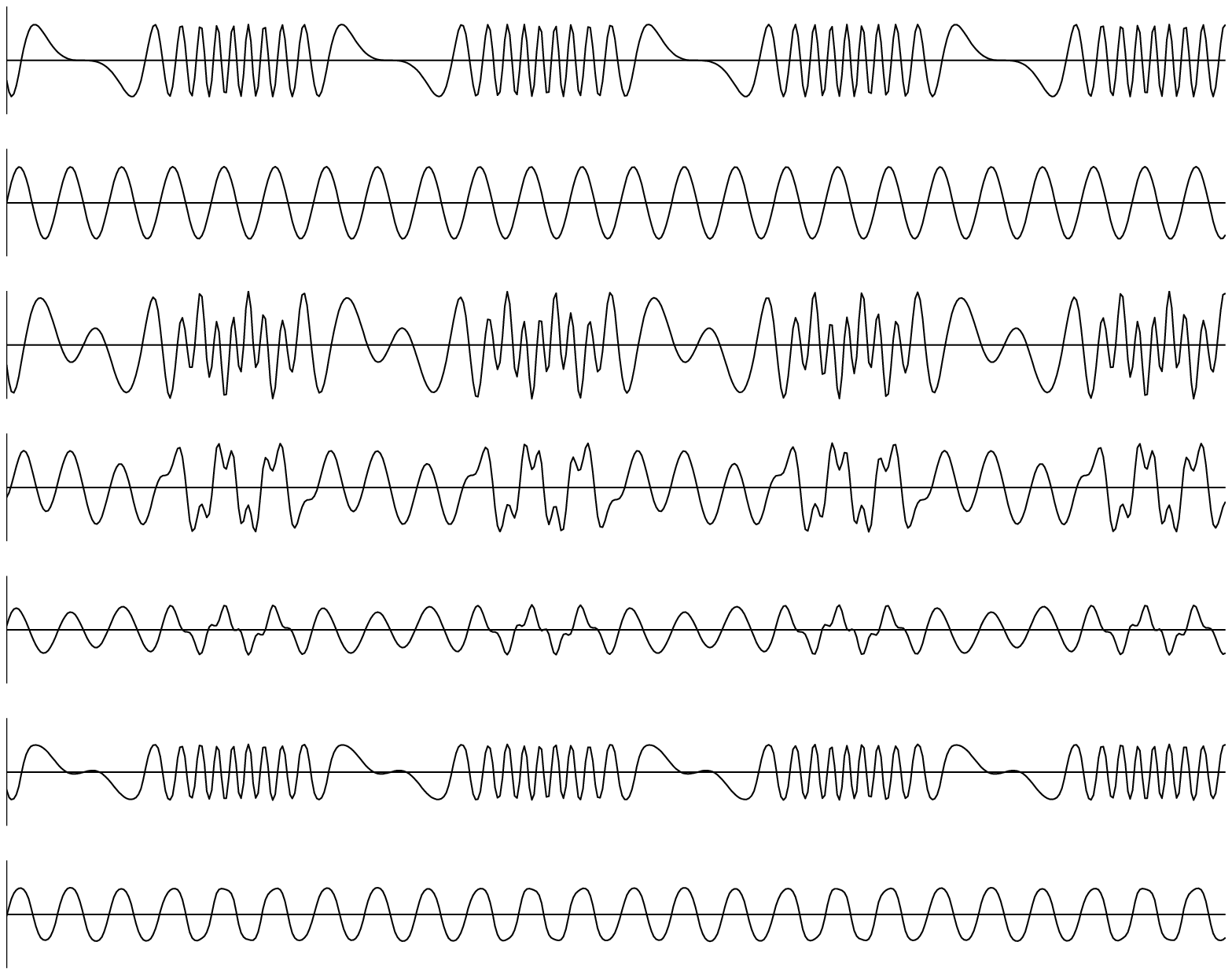} 
\etabu
\etabu
\\ {\small Fig.~7: Results of the source separation in Example 3.}
\ecc

\bcc
\btabu{@{}c@{}}
\epsfxsize=120mm\epsfysize=40mm
\epsfbox{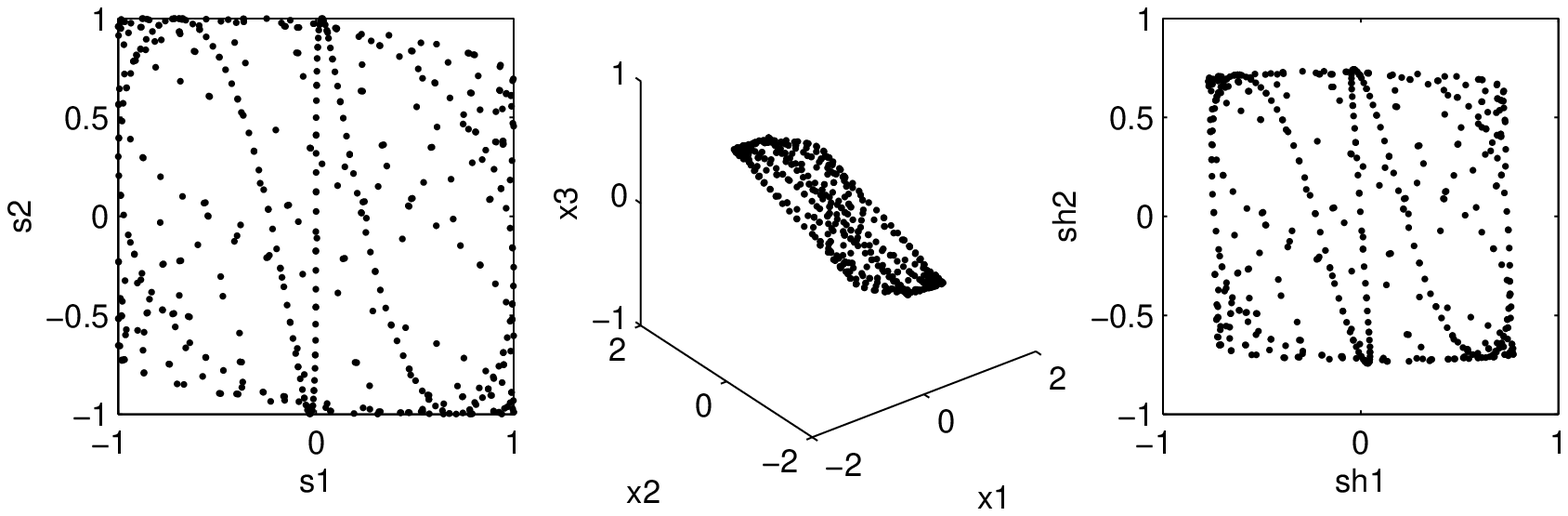} 
\etabu
\\ {\small Fig.~8: Results of the source separation in Example 3: 
phase space distribution of sources, mixed signals and separated sources}
\ecc

\bcc
\btabu{@{}c@{}}
\epsfxsize=120mm\epsfysize=90mm
\epsfbox{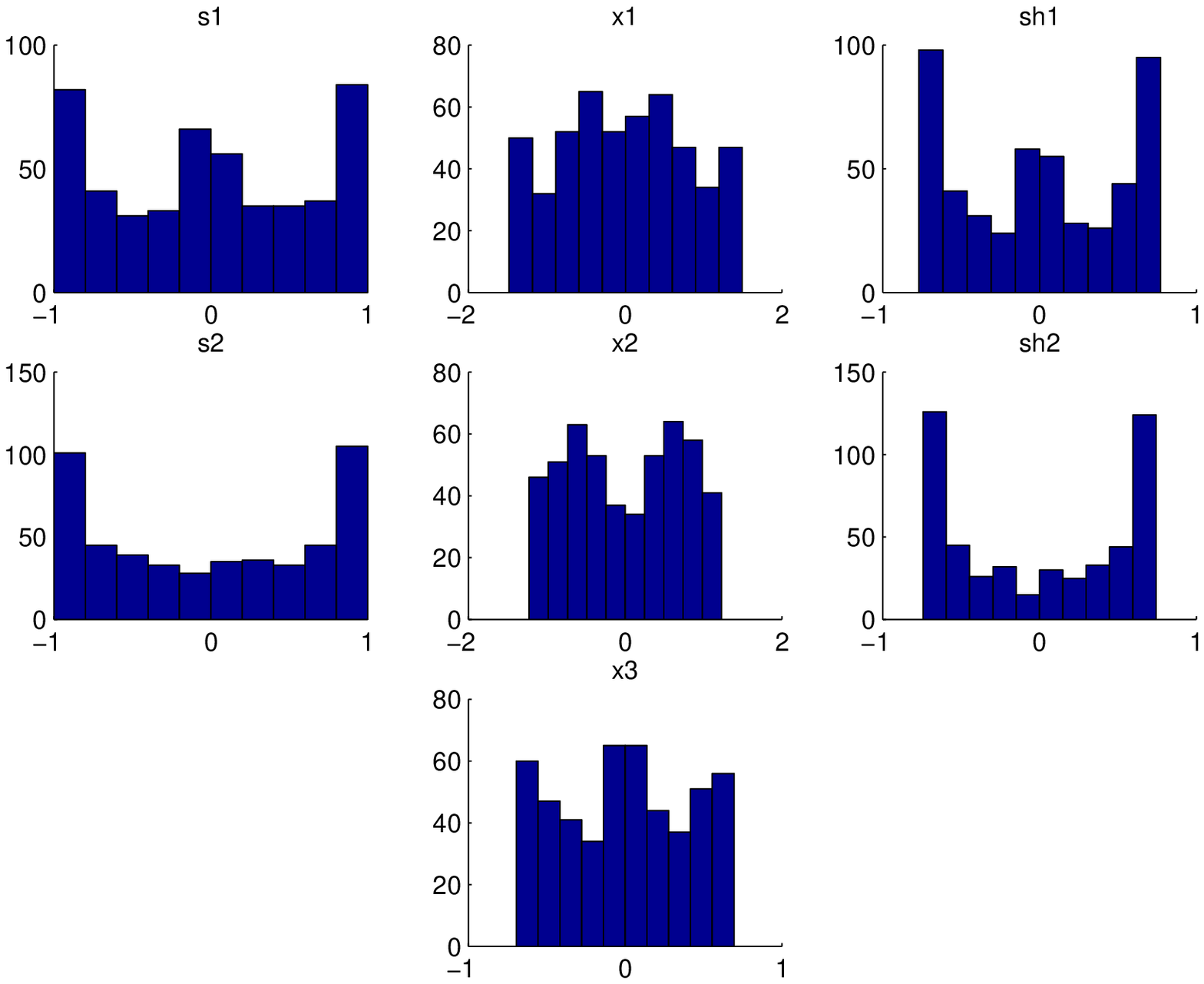} 
\etabu
\\ {\small Fig.~9: Results of the source separation in Example 3: 
histograms of sources, mixed signals and separated sources}
\ecc

\bigskip
\subsection{Example 4}
Hier, we considered the three sources of the Exammple 2 and simulated 
the case where there are only two receivers using the mixing matrice    
\[
\Ab=
\pmatrix{
1. & .2 & 1 \cr 
-.5 & 1. & .2 }
\]
to obtain the two set of data $x_1(t)$ and $x_2(t)$. 
Then we applied again the algorithm given in (\ref{UsedAlg}). 
The following figures show the obtained results. 

\bcc
\btabu[b]{@{}cc@{}}
\btabu[b]{@{}c@{}}
$\left\{\barr{@{}l@{}}
s_1(t) \\ ~\\ ~\\  
s_2(t) \\ ~\\ ~\\  
s_3(t) 
\earr\right.$
\\ ~\\ ~\\  
$\left\{\barr{@{}l@{}}
x_1(t) \\ ~\\ ~\\  
x_2(t) 
\earr\right.$
\\ ~\\ ~\\  
$\left\{\barr{@{}l@{}}
\wh{s}_1(t) \\  ~\\ ~\\  
\wh{s}_2(t) \\  ~\\ ~\\  
\wh{s}_3(t) 
\earr\right.$
\\ ~\\  
\etabu
&
\btabu[b]{@{}c@{}}
\epsfxsize=100mm\epsfysize=100mm
\epsfbox{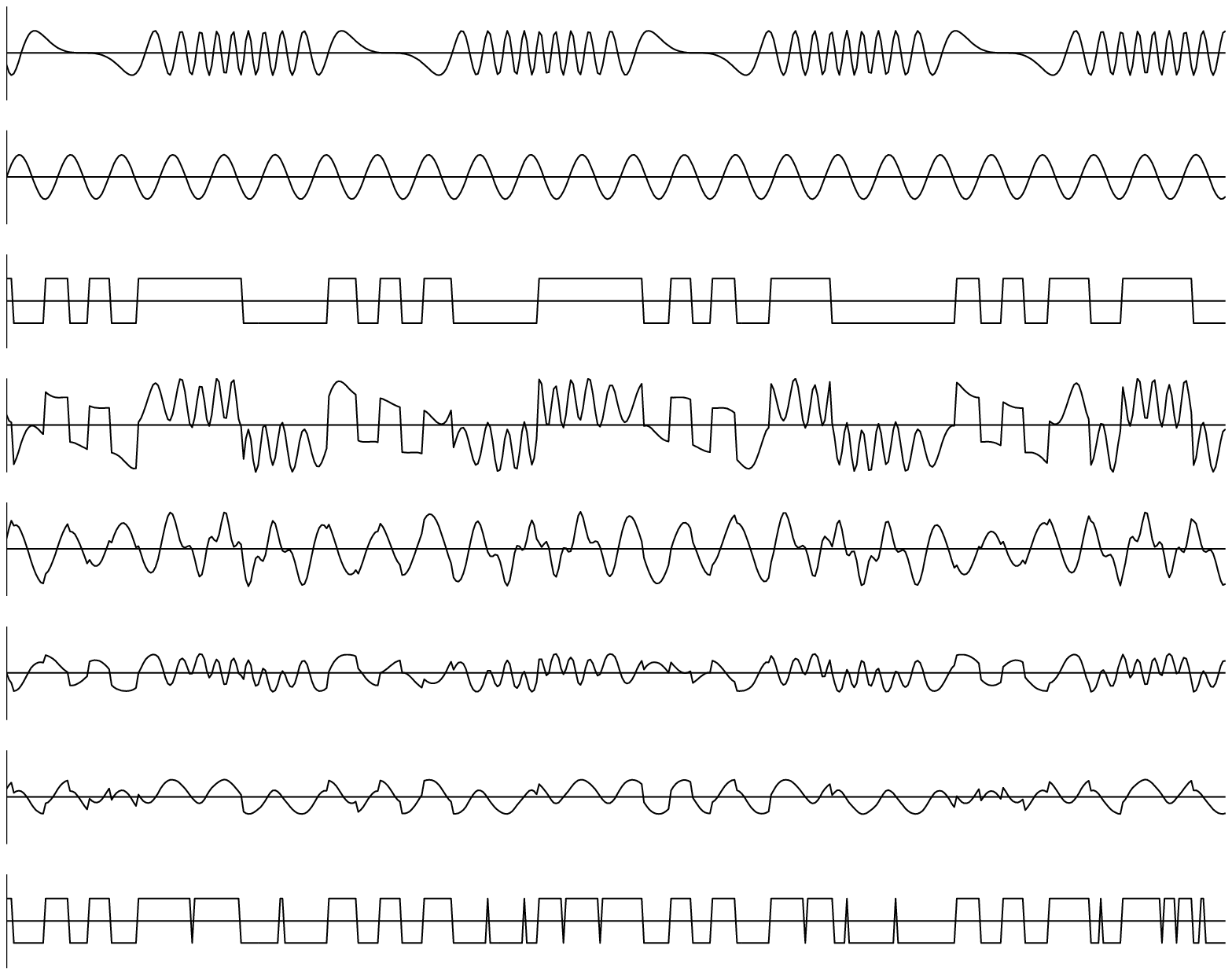} 
\etabu
\etabu
\\ {\small Fig.~7: Results of the source separation in Example 4.}
\ecc

\bcc
\btabu{@{}c@{}}
\epsfxsize=120mm\epsfysize=40mm
\epsfbox{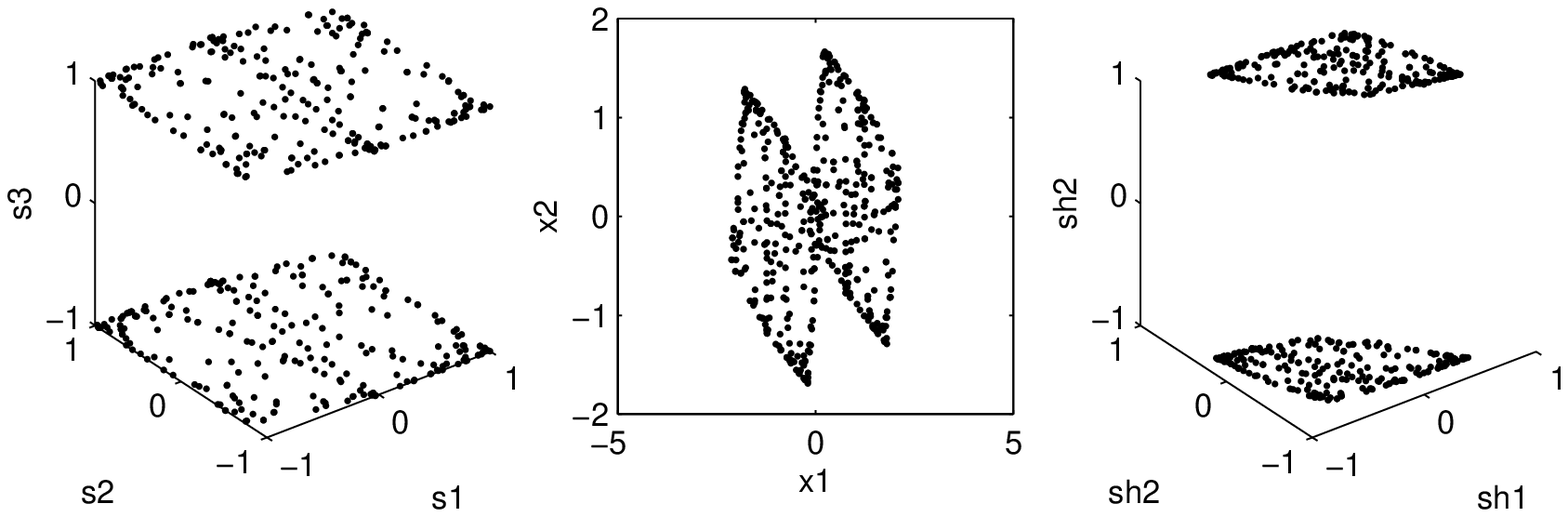} 
\etabu
\\ {\small Fig.~8: Results of the source separation in Example 4: 
phase space distribution of sources, mixed signals and separated sources}
\ecc

\bcc
\btabu{@{}c@{}}
\epsfxsize=120mm\epsfysize=90mm
\epsfbox{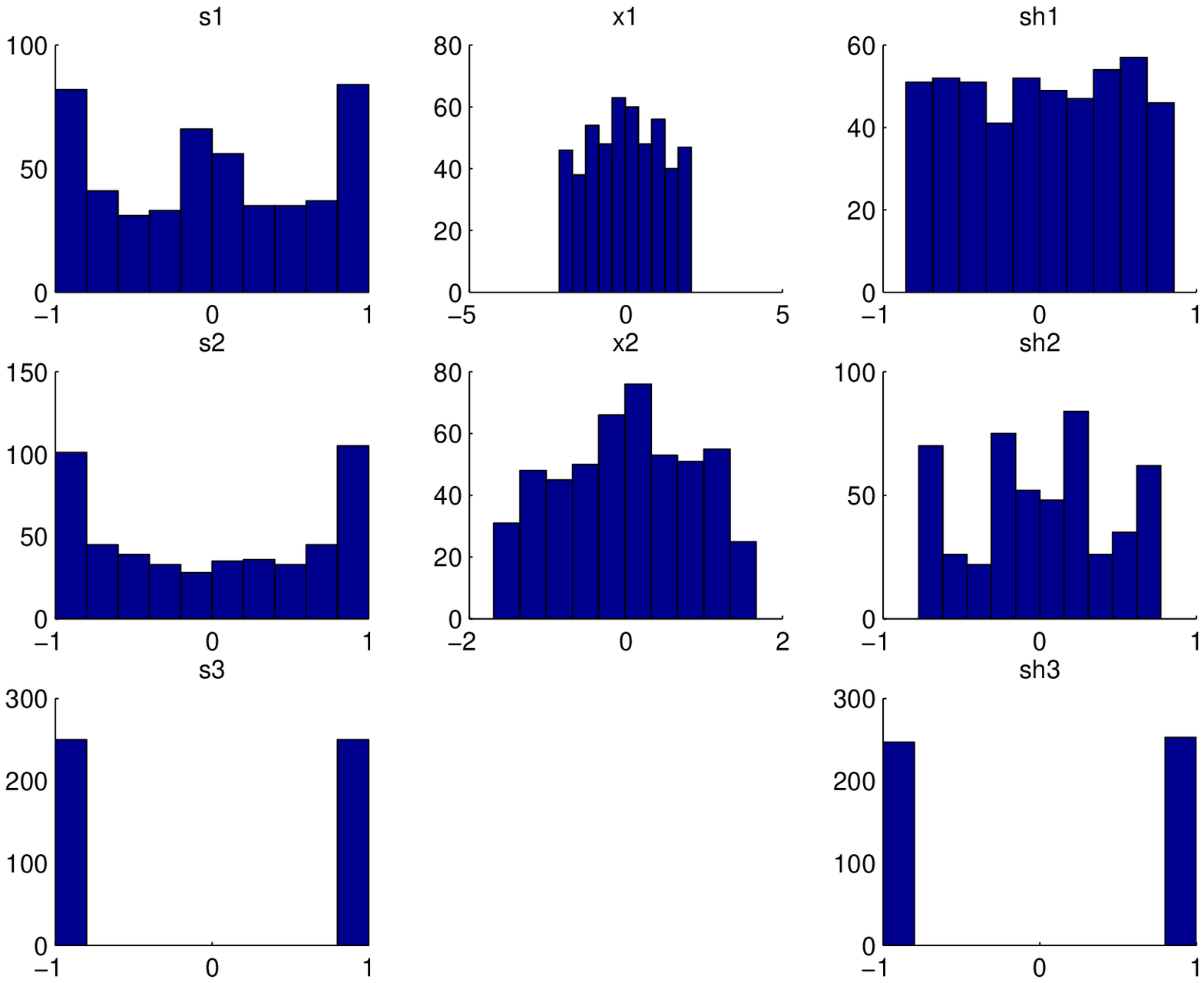} 
\etabu
\\ {\small Fig.~9: Results of the source separation in Example 4: 
histograms of sources, mixed signals and separated sources}
\ecc

\section{Conclusions}
We investigated the use of the Bayesian estimation theory to source 
separation and showed that this approach has the potential to push 
farther the limits of the classical methods. 
This work is not really yet finished. We are going now 
to compare the performances of the proposed methods to other 
conventional ones on simulated and real data. 

\def\UP#1{\uppercase{#1}}
\def\XS{\xspace}
\def\bibdir{/users/bertha/seismic/seismic/TeX/}
\def\CGVIP{Computer Graphics and Visual Image Processing}
%
\def\jan{jan. }
\def\feb{f\'ev. }
\def\mar{mars }
\def\apr{avr. }
\def\may{mai }
\def\jun{juin }
\def\jul{jui. }
\def\aug{ao\^ut }
\def\sep{sep. }
\def\oct{oct. }
\def\nov{nov. }
\def\dec{d\'ec. }
\def\Jan{Jan. }	
\def\Feb{Feb. }
\def\Mar{Mar. }
\def\Apr{Apr. }
\def\May{May }
\def\Jun{June }
\def\Jul{July }
\def\Aug{Aug. }
\def\Sep{Sep. }
\def\Oct{Oct. }
\def\Nov{Nov. }
\def\Dec{Dec. }
\def\sub{soumis \`a }

%
\def\jan{janvier\xspace}
\def\feb{f\'evrier\xspace}
\def\mar{mars\xspace}
\def\apr{avril\xspace}
\def\may{mai\xspace}
\def\jun{juin\xspace}
\def\jul{juillet\xspace}
\def\aug{ao\^ut\xspace}
\def\sep{septembre\xspace}
\def\oct{octobre\xspace}
\def\nov{novembre\xspace}
\def\dec{d\'ecembre\xspace}
\def\Jan{January\xspace}	
\def\Feb{February\xspace}
\def\Mar{March\xspace}
\def\Apr{April\xspace}
\def\May{May\xspace}
\def\Jun{June\xspace}
\def\Jul{July\xspace}
\def\Aug{August\xspace}
\def\Sep{September\xspace}
\def\Oct{October\xspace}
\def\Nov{November\xspace}
\def\Dec{December\xspace}
\def\sub{soumis \`a\xspace}
%
%

\def\AsAs{Astrononmy and Astrophysics}					
\def\AAP{Advances in Applied Probability}				
\def\ABE{Annals of Biomedical Engineering}				
\def\AISM{Annals of Institute of Statistical Mathematics}		
\def\AMS{Annals of Mathematical Statistics}			
\def\AO{Applied Optics}							
\def\AP{The Annals of Probability}					
\def\ARAA{Annual Review of Astronomy and Astrophysics}			
\def\AST{The Annals of Statistics}					
\def\AT{Annales des T\'el\'ecommunications}				
\def\BMC{Biometrics}							
\def\BMK{Biometrika}							
\def\CPAM{Communications on Pure and Applied Mathematics}		
\def\EMK{Econometrica}							
\def\CRAS{Compte-rendus de l'acad\'emie des sciences}			
\def\CVGIP{Computer Vision and Graphics and Image Processing}		
\def\GJRAS{Geophysical Journal of the Royal Astrononomical Society}	
\def\GSC{Geoscience}						
\def\GPH{Geophysics}							
\def\GRETSI#1{Actes du #1$^{\mbox{e}}$ Colloque GRETSI} 		
\def\CGIP{Computer Graphics and Image Processing}			
\def\ICASSP{Proceedings of IEEE ICASSP}					
\def\ICEMBS{Proceedings of IEEE EMBS}					
\def\ICIP{Proceedings of the International Conference on Image Processing}
\def\ieeP{Proceedings of the IEE}					
\def\ieeeAC{IEEE Transactions on Automatic and Control}			
\def\ieeeAES{IEEE Transactions on Aerospace and Electronic Systems}	
\def\ieeeAP{IEEE Transactions on Antennas and Propagation}		
\def\ieeeAPm{IEEE Antennas and Propagation Magazine}			
\def\ieeeASSP{IEEE Transactions on Acoustics Speech and Signal Processing}
\def\ieeeBME{IEEE Transactions on Biomedical Engineering}		
\def\ieeeCS{IEEE Transactions on Circuits and Systems}			
\def\ieeeCT{IEEE Transactions on Circuit Theory}			
\def\ieeeC{IEEE Transactions on Communications}				
\def\ieeeGE{IEEE Transactions on Geoscience and Remote Sensing}		
\def\ieeeGEE{IEEE Transactions on Geosciences Electronics}		
\def\ieeeIP{IEEE Transactions on Image Processing}			
\def\ieeeIT{IEEE Transactions on Information Theory}			
\def\ieeeMI{IEEE Transactions on Medical Imaging}			
\def\ieeeMTT{IEEE Transactions on Microwave Theory and Technology}	
\def\ieeeM{IEEE Transactions on Magnetics}				
\def\ieeeNS{IEEE Transactions on Nuclear Sciences}			
\def\ieeePAMI{IEEE Transactions on Pattern Analysis and Machine Intelligence}
\def\ieeeP{Proceedings of the IEEE}					
\def\ieeeRS{IEEE Transactions on Radio Science}				
\def\ieeeSMC{IEEE Transactions on Systems, Man and Cybernetics}		
\def\ieeeSP{IEEE Transactions on Signal Processing}			
\def\ieeeSSC{IEEE Transactions on Systems Science and Cybernetics}	
\def\ieeeSU{IEEE Transactions on Sonics and Ultrasonics}		
\def\ieeeUFFC{IEEE Transactions on Ultrasonics Ferroelectrics and Frequency Control}
\def\IJC{International Journal of Control}				
\def\IJCV{International Journal of Computer Vision}			
\def\IJIST{International Journal of Imaging Systems and Technology}	
\def\IP{Inverse Problems}						
\def\ISR{International Statistical Review}				
\def\IUSS{Proceedings of International Ultrasonics Symposium}		
\def\JAPH{Journal of Applied Physics}					
\def\JAP{Journal of Applied Probability}				
\def\JAS{Journal of Applied Statistics}					
\def\JASA{Journal of Acoustical Society America}			
\def\JASAS{Journal of American Statistical Association}			
\def\JBME{Journal of Biomedical Engineering}				
\def\JCAM{Journal of Computational and Applied Mathematics}		
\def\JEWA{Journal of Electromagnetic Waves and Applications}		
\def\JMO{Journal of Modern Optics}					
\def\JNDE{Journal of Nondestructive Evaluation}				
\def\JMP{Journal of Mathematical Physics}				
\def\JOSA{Journal of the Optical Society of America}			
\def\JP{Journal de Physique}						
\def\JRSSA{Journal of the Royal Statistical Society A}			
\def\JRSSB{Journal of the Royal Statistical Society B}			
\def\JRSSC{Journal of the Royal Statistical Society C}			
\def\JSPI{Journal of Statistical Planning and Inference}  		
\def\JTSA{Journal of Time Series Analysis}                   		
\def\JVCIR{Journal of Visual Communication and Image Representation} 	
	\def\MMAS{???} 
\def\KAP{Kluwer \uppercase{A}cademic \uppercase{P}ublishers}							%
\def\MNAS{Mathematical Methods in Applied Science}			
\def\MNRAS{Monthly Notices of the Royal Astronomical Society}		
\def\MP{Mathematical Programming}					
	\def\NSIP{NSIP}  
\def\OC{Optics Communication}						
\def\PRA{Physical Review A}						
\def\PRB{Physical Review B}						
\def\PRC{Physical Review C}						
\def\PRD{Physical Review D}						
\def\PRL{Physical Review Letters}					
\def\RGSP{Review of Geophysics and Space Physics}			
\def\RPA{Revue de Physique Appliqu\'e}							
\def\RS{Radio Science}							
\def\SP{Signal Processing}						
\def\siamAM{SIAM Journal of Applied Mathematics}			
\def\siamCO{SIAM Journal of Control}					
\def\siamJO{SIAM Journal of Optimization}				
\def\siamMA{SIAM Journal of Mathematical Analysis}			
\def\siamNA{SIAM Journal of Numerical Analysis}				
\def\siamR{SIAM Review}							
\def\SSR{Stochastics and Stochastics Reports}       			
\def\TPA{Theory of Probability and its Applications}			
\def\TMK{Technometrics}							
\def\TS{Traitement du Signal}						
\def\UCMMP{U.S.S.R. Computational Mathematics and Mathematical Physics}	
\def\UMB{Ultrasound in Medecine and Biology}				
\def\US{Ultrasonics}							
\def\USI{Ultrasonic Imaging}						

\bibliographystyle{ieeetr}
\bibliography{gpibase,gpipubli,amd,amd97,ss,cardozo,mackay,ica99}
\end{document}